\documentclass{article}
\usepackage{subfig}
\usepackage{graphicx}
\usepackage{float}
\pdfoutput=1

\newcommand{\midand}{\ \mathrm{and}\ }
\newcommand{\lapl}{\nabla^2}

\begin{document}

{\Large \noindent \textbf{Atom Based Grain Extraction and Measurement of\\Geometric Properties}}\\

\hspace{15pt} {\raggedright \noindent \textbf{Gabriel Martine La Boissoni\`ere, Rustum Choksi}}\\

{\hspace{15pt} \noindent Department of Mathematics and Statistics, McGill University

\hspace{15pt} 805 Rue Sherbrooke O, Montr\'eal, QC H3A 0B9, Canada}\\

\hspace{15pt} {\noindent E-mail: \textit{gabriel.martine-laboissoniere@mail.mcgill.ca}, \textit{rchoksi@math.mcgill.ca}}

\vspace{15pt}
\noindent \textbf{Abstract}: We introduce an accurate, self-contained and automatic atom based numerical algorithm to characterize grain distributions in two dimensional Phase Field Crystal simulations. Four input parameters must be set by the user and their effect is described. We compare the method with hand segmented and known test grain distributions to show that the algorithm is able to extract grains and measure their area, perimeter and other geometric properties with high accuracy. We also compare the proposed method to a simpler but less accurate grid based approach. This method is currently tuned to extract data from Phase Field Crystal simulations in the hexagonal lattice regime but the framework may be extended to more general problems.

\vspace{15pt}

\newpage

\section{Introduction}
\begin{figure}[H]
	\centering
	\subfloat{\includegraphics[scale = 0.5]{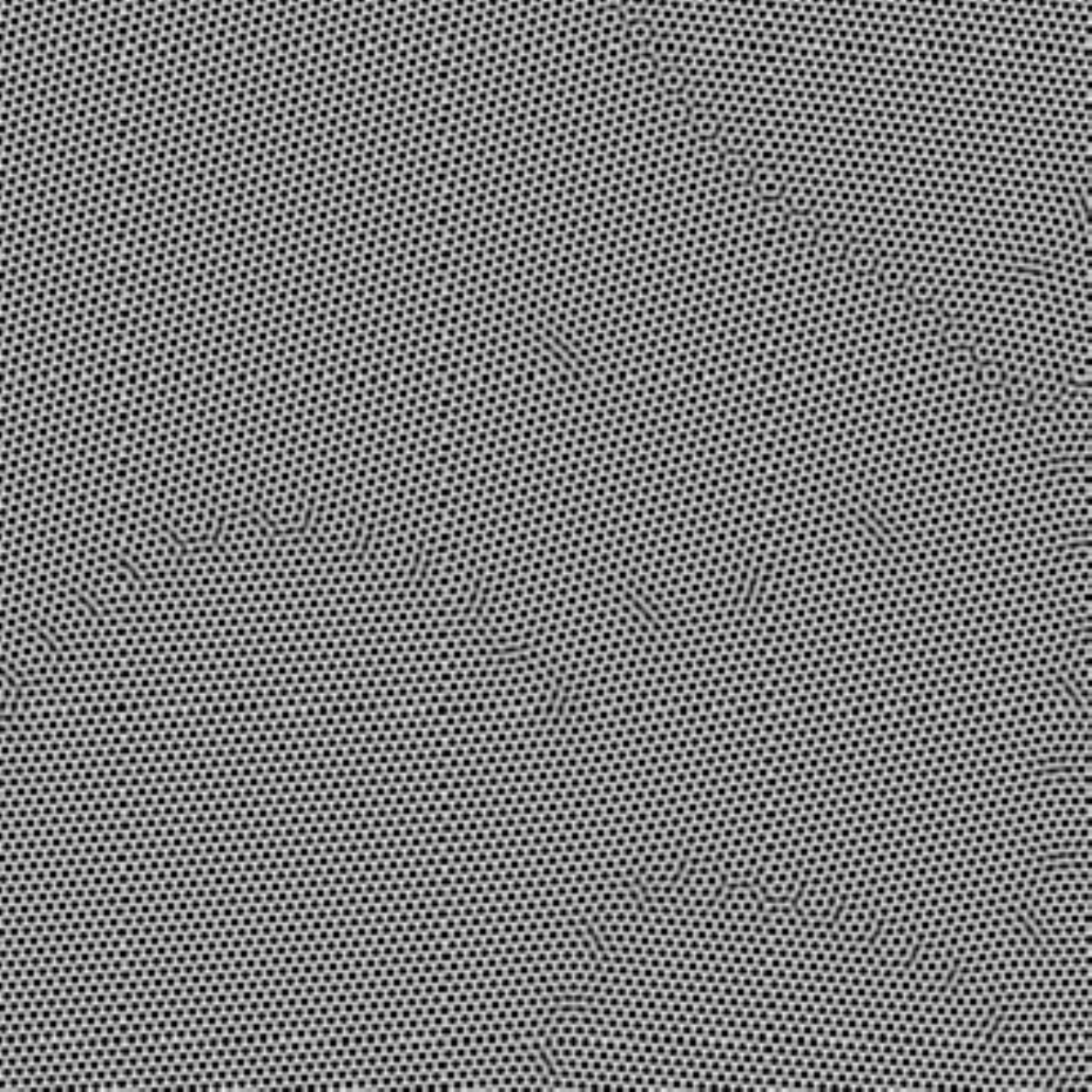}}
	\caption{Detail of atomic positions in simple PFC simulations. Each atom is represented as a ``bump'' in luminosity. Away from grain boundaries, these bumps are uniform through space and time and arrange into regular hexagonal lattices.}
	\label{fig:PFC_Atoms}
\end{figure}

The polycrystalline structure of materials is one of the main deciding factors in determining physical properties such as hardness and ductility. A thorough understanding of the conditions that lead to the formation of a desirable crystalline structure is essential in many technological applications. To study such conditions and their impact on the evolution of polycrystalline materials, it is necessary to obtain experimental data throughout the structure forming processes. For example, extensive studies into the evolution of grains and stagnation in thin Al and Cu metallic films have been carried out in \cite{BARMAK_EarsTails}. Such investigations are experimentally challenging so simulations are desirable to investigate the evolution of crystals. Many theoretical models have been proposed to do so, for example the well known Mullins model \cite{MULLINS_Mullins}. More recently, the Phase Field Crystal (PFC) \cite{ELDER_Elasticity, EMMERICH_PFCReview} model has been successful in modelling a wide variety of crystalline phenomena. In particular, grain size distributions predicted by the 2d PFC model agree remarkably well with experimental data \cite{BACKOFEN_GSD}. While these results are encouraging, to our knowledge a thorough statistical assessment of geometric properties of the grain boundaries for the PFC model has yet to be performed. To do so, one needs a verifiably accurate method for extracting geometric features of grain boundaries from input data containing millions of atoms (cf. figure \ref{fig:PFC_Atoms}).

In this article, we present and assess a simple and self-contained method specifically adapted to characterize the geometric properties of hexagonal lattice grain distributions extracted from 2d PFC simulations. Attempts to numerically characterize such grain networks are not new. Much experimental data comes from digital images of thin metallic films or tomographic maps of 3d crystals in which grains are very large compared to atoms. This typically leads to very sharp and clear grain boundaries that can be recognized readily both manually or numerically, as in \cite{HEILBRONNER_Automatic, BERGER_QuantitativeAnalysis}. Grain area can then be measured by counting the number of pixels within each connected region. Identifying the boundary network obtained from atomistic simulations proves to be a much more complicated task. Unlike images where orientation is essentially a piecewise constant field in space, the orientation in atomistic simulations must first be extracted from atomic positions. Such an orientation field may be extracted using variational techniques as in \cite{SINGER_Variational, ELSEY_ImageAnalysis, YANG_SST}. Alternatively, the field may be obtained by projecting and interpolating geometrically computed local orientations, obtained for example with the atomistic visualization tool OVITO \cite{OVITO_REF} (see \cite{STUKOWSKI_OVITO} for details on specific approaches). In both cases, the gradient of the orientation field may then be thresholded to create a ``skeleton'' of grain boundaries that can be refined by hand, as in \cite{BACKOFEN_GSD, BACKOFEN_PRIVATE}. Grain area may then be characterized by counting connected pixels. Recently, an automated atomistic technique to identify grains whose barycenter can be tracked in time has been developed in \cite{PANZARINO_Tracking}. 

We propose a grain extraction and measurement procedure which is also based solely upon atomic positions instead of working with an orientation field projected onto a discrete grid. We show that our method is capable of measuring grain area, perimeter, Grain Boundary Character Distributions (GBCD) \cite{HOLM_GBCD} and related geometric properties with good accuracy. Our approach uses the Voronoi region of each atom to characterize defects and grain boundaries, creating a stencil with which one identifies grains on the graph of atoms. Once atoms belonging to a grain have been identified, geometric properties may be computed. All steps rely on very simple geometric routines that can be implemented efficiently. For simplicity and because our main goal is to extract grain distributions arising from the basic PFC equation \cite{ELDER_Elasticity}, our algorithm is adapted to characterize a patchwork of 2d hexagonal lattices with interatomic distance $d$. Geometric property extraction may be extended to more interesting scenarios once the underlying atomistic characterization is extended to 3d and to other lattice types as in \cite{PANZARINO_Tracking}.

The outline of the article is as follows. First, we describe the general process of computing grain properties from atomistic or phase field simulations. We then detail our atom based data extraction procedure. Next, we validate the accuracy of both the numerically computed grain network and of the extracted properties by comparing the numerical segmentation with hand segmentations and artificially constructed grain distributions. We then present how our atom based procedure outperforms a simple grid based extraction method. Finally, we present a few sample results from PFC simulations.

\section{The Grain Extraction and Measurement Problem}
\label{sub:GrainExtractionProblem}
\begin{figure}[H]
	\centering
	\subfloat{\includegraphics[scale = 0.6]{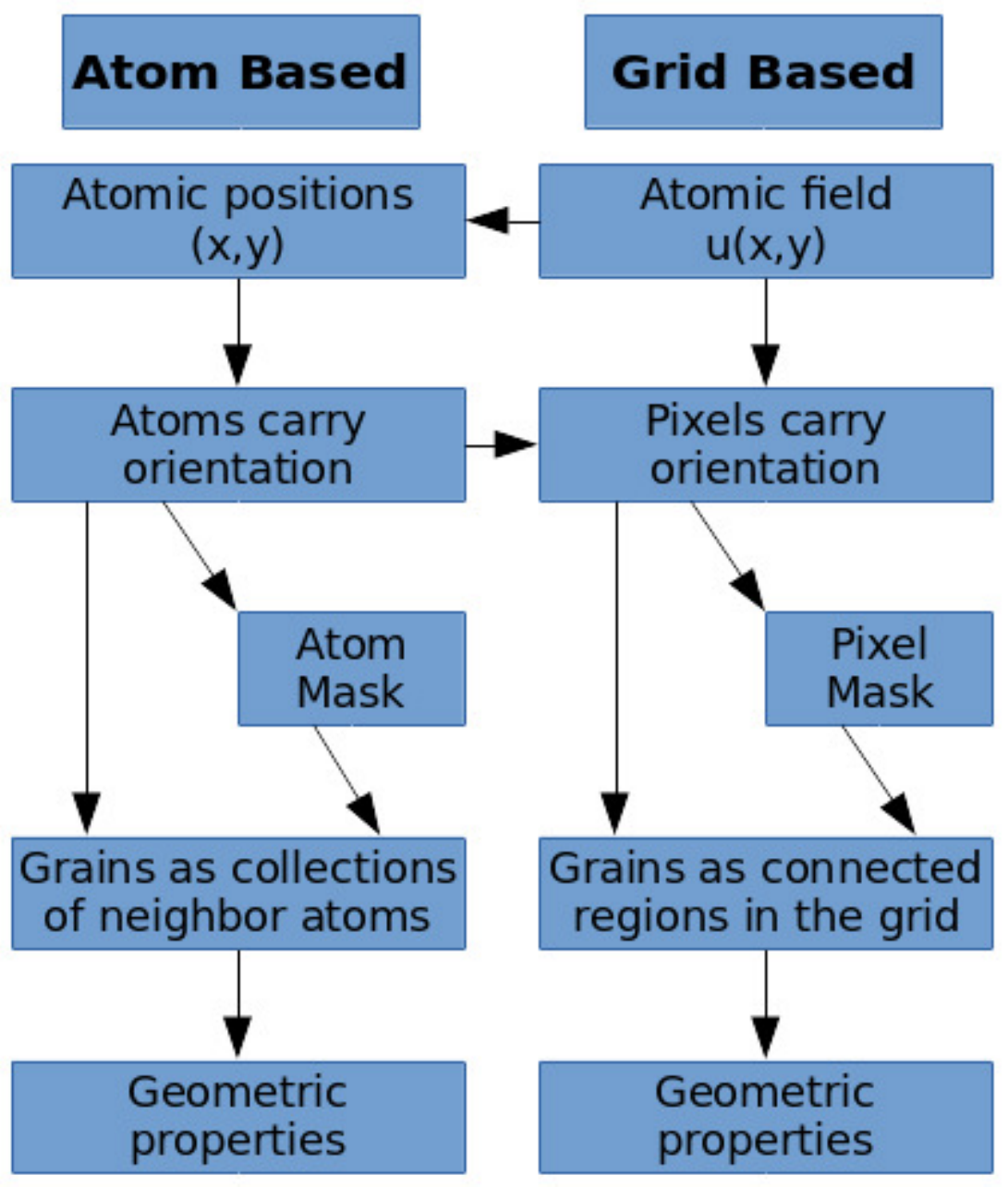}}
	\caption{Outline of the atom and grid based numerical grain extraction methods. The proposed method extracts atomic positions from an input phase field then follows the left column, preserving atomic positional data.}
	\label{fig:DiagramPicture}
\end{figure}

Suppose one is given a distribution of atomic positions. Existing software is capable of assigning a local lattice orientation to each atom using geometrical techniques. On the other hand, one may also be given an image of atomic ``densities''. In this case, variational techniques can be used as ``filters'' to convert the image into an image of lattice orientations. Once orientation information is known, grains may be extracted by constructing regions separated by appropriately defined grains boundaries. Such regions may then be measured in order to extract grain properties such as area, perimeter and orientation. While the nature of the data is very different, there is a natural parallel between the two approaches illustrated in figure \ref{fig:DiagramPicture}.
If one follows the left side, atomistic simulation data is handled while preserving the underlying atomic structure. Similarly on the right, one uses variational methods and grid based algorithms to handle grid based phase field data. The step to convert the phase field $u$ into an orientation map may be implemented using one of the variational methods found in \cite{ELSEY_ImageAnalysis} or \cite{YANG_SST} for example.

As indicated by the horizontal arrows, it is possible to ``cross over'' from one side to the other: first, atomic positions may be extracted directly from a given image (detailed in \ref{sub:MethodAtomRec}) while orientations known at atomic positions may be projected onto a discrete grid, forming an image of orientations. This is simply implemented by assigning to a pixel the orientation of the atomic positions it contains. Thus, any given input data may be treated using either of the two different extraction and measurement procedures after crossing over once or twice if necessary.

The grid based extraction and measurement is implemented with the following numerical program. Given an image of orientations $v$ on a grid,
\begin{itemize}
	\item Compute a pixel mask of grain boundaries by thresholding an appropriate quantity such as $|\nabla v|^2$.
	\item Compute the connected components of non-boundary pixels in the mask.
	\item Compute grain angles by taking the average of the image over the corresponding connected component.
	\item Assign masked pixels to the nearest grain.
	\item Measure grain properties using grid based methods.
\end{itemize}
In the last step, the area of a grain corresponds to its number of pixels while its perimeter can be computed by tracing the border of the region and summing the distance between adjacent pixels. We note that in addition to the input image $v$, potential pixel masks may also be obtained from variational methods, for example, a distortion map as in \cite{ELSEY_ImageAnalysis}.

The atom based algorithm is quite similar but all steps are implemented directly on the graph of atomic positions with no need to project or interpolate the orientation map. All tasks now involve the geometric treatment of exact atomic positions thus improving the accuracy of the measured geometric quantities. For this reason, especially to analyze PFC simulations, it is desirable to extract atomic positions from the phase field, compute orientations geometrically using techniques found in \cite{STUKOWSKI_OVITO} and then follow the atom based program. We shall show that this process results in more accurate grain detection and measurement and also bypasses the need to process the phase field using costly variational techniques.

\section{Atom Based Grain Extraction and Measurement}
We now detail the atom based grain extraction procedure adapted to 2d hexagonal lattices. We assume the input data is a PFC phase field in which atoms have had sufficient time to arrange into reasonably large lattices. For clarity, we give an outline of the full procedure:
\begin{itemize}
	\item Atomic positions are extracted. (\ref{sub:MethodAtomRec})
	\item A lattice orientation is assigned to each atom. (\ref{sub:VoronoiProperty})
	\item The local lattice structure of each atom is probed to determine which atoms are close to grain boundaries, forming a grain boundary mask. (\ref{sub:VoronoiProperty})
	\item Grains are identified in a fashion similar to a flood-fill algorithm. (\ref{sub:FloodFill})
	\item Atoms in the grain boundary mask are assigned to the closest grain. (\ref{sub:FloodFill})
	\item Grain properties are measured. (\ref{sub:Measurement})
	\item Spurious grains are detected and removed. (\ref{sub:PostProcessing})
\end{itemize}

A MATLAB implementation of our procedure is capable of analyzing an $8192^2$ image containing roughly a million atoms in about 6 minutes on a standard machine.

Note that our input data lies on a periodic domain; it is therefore necessary for the implementation to allow seamless identifications between opposing boundaries. This refinement will not be detailed for brevity. Also, average orientations are often required, a possibly ill defined quantity given the identification between $0^\circ = 60^\circ$ in a hexagonal lattice. To compute an isotropic average, convert a set of orientations into position vectors on the unit circle such that $(1,0)$ corresponds to both $0^\circ \midand 60^\circ$. Then, the angle of the average \textit{vector} can be mapped on the same range and taken as the average orientation. This agrees with the Euclidean average for narrow distributions but avoids any ambiguity near the identification.

\subsection{Atom Recognition}
\label{sub:MethodAtomRec}
Atoms are recognized by finding the center of bumps illustrated in figure \ref{fig:PFC_Atoms}. Given an input black and white image where atoms are circular bumps in luminosity, we first use a contouring algorithm to find contours at a given level $h$. A circle is then fitted to every contour, arrays of points corresponding to each bump. As long as the points are well distributed along the circle, one can find a circle with position and radius minimizing the least-squares distance to the contour points. The appropriate level $h$ to be used must be chosen so that all atoms are represented as the largest circles possible without introducing errors near boundaries, where atoms are deformed. The range of appropriate $h$ is broad making the procedure straightforward.

\subsection{Computation of Atomic Voronoi Properties and Grain Boundaries}
\label{sub:VoronoiProperty}
\begin{figure}[H]
	\centering
	\subfloat{\includegraphics[scale = 0.5]{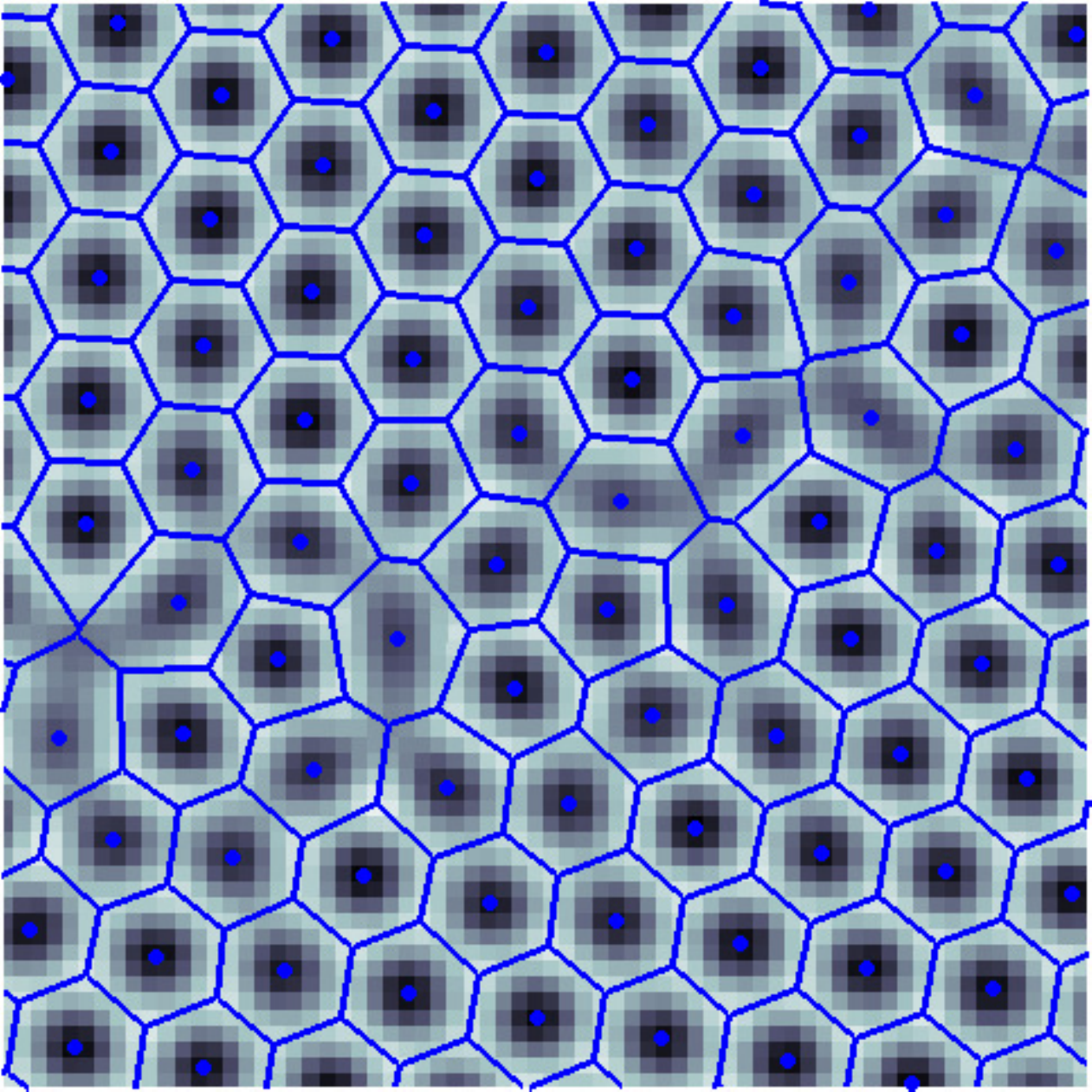}}
	\caption{Detail of a PFC phase field superposed with the extracted atomic positions and their Voronoi tessellation in blue. Note how a grain boundary is clearly visible and can be identified to the set of Voronoi cells that are not perfect hexagons.}
	\label{fig:VoronoiTesselation}
\end{figure}

We now identify defects and atoms with a local environment that is far from the expected lattice configuration to trace a preliminary boundary network while other atoms will be assumed to be part of a grain. Our approach is to compute the Voronoi tessellation \cite{OKABE_VoronoiDelaunay} of all atomic positions and determine how closely the Voronoi or Wigner-Seitz cell of an atom matches a regular hexagon, see figure \ref{fig:VoronoiTesselation} for an illustration. If the cell is hexagonal, we may also compute the local lattice orientation of an atom by averaging the angle of the vector pointing to the six Voronoi vertices modulo the $60^\circ$ symmetry. Near boundaries, this computation will give inaccurate results. When enough neighbor grains have similar orientations, they form a coherent lattice and will be considered to form a grain.

\begin{figure}[H]
	\centering
	\subfloat[]{\includegraphics[width=0.3\textwidth]{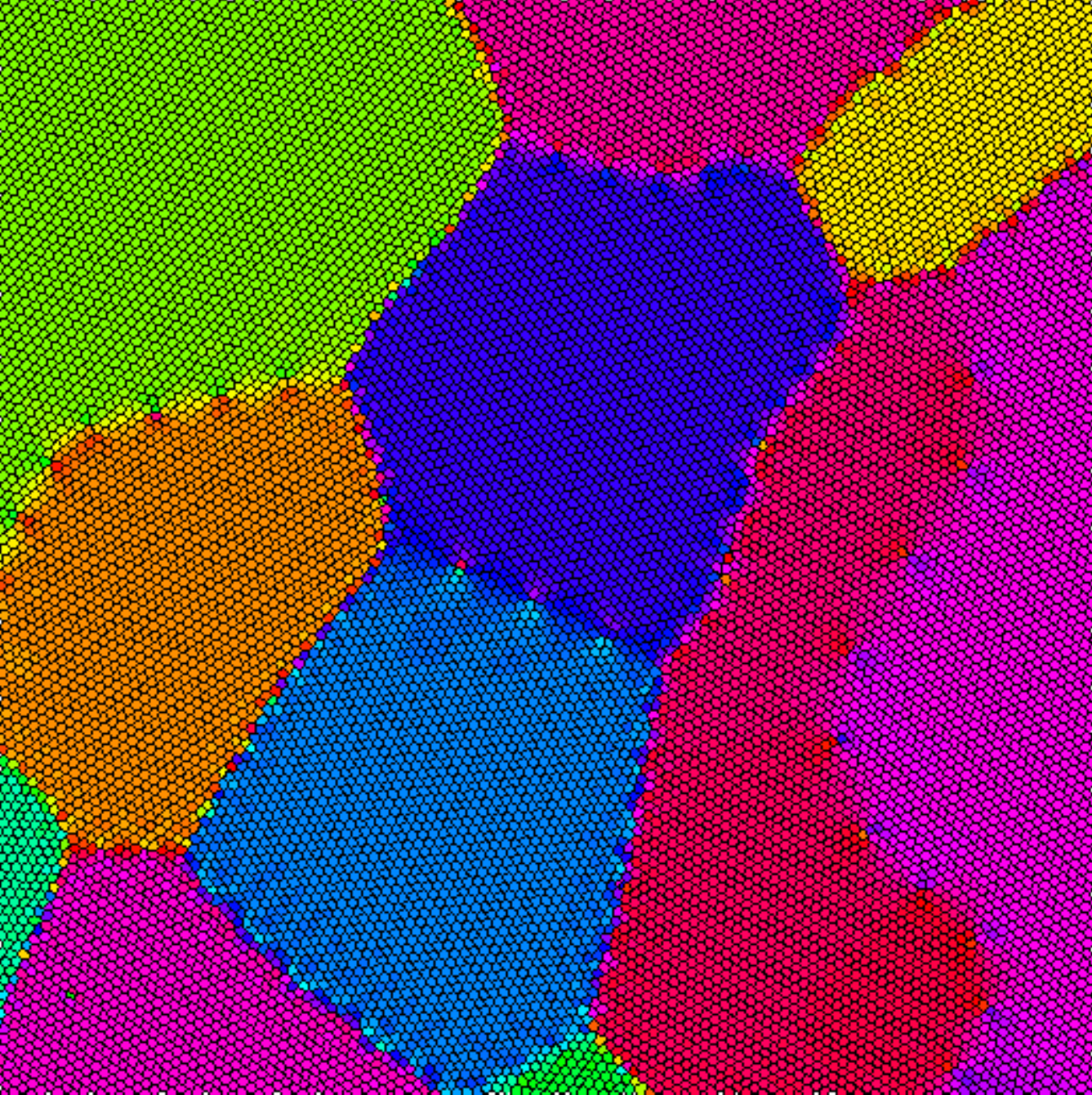}}\enskip
	\subfloat[]{\includegraphics[width=0.3\textwidth]{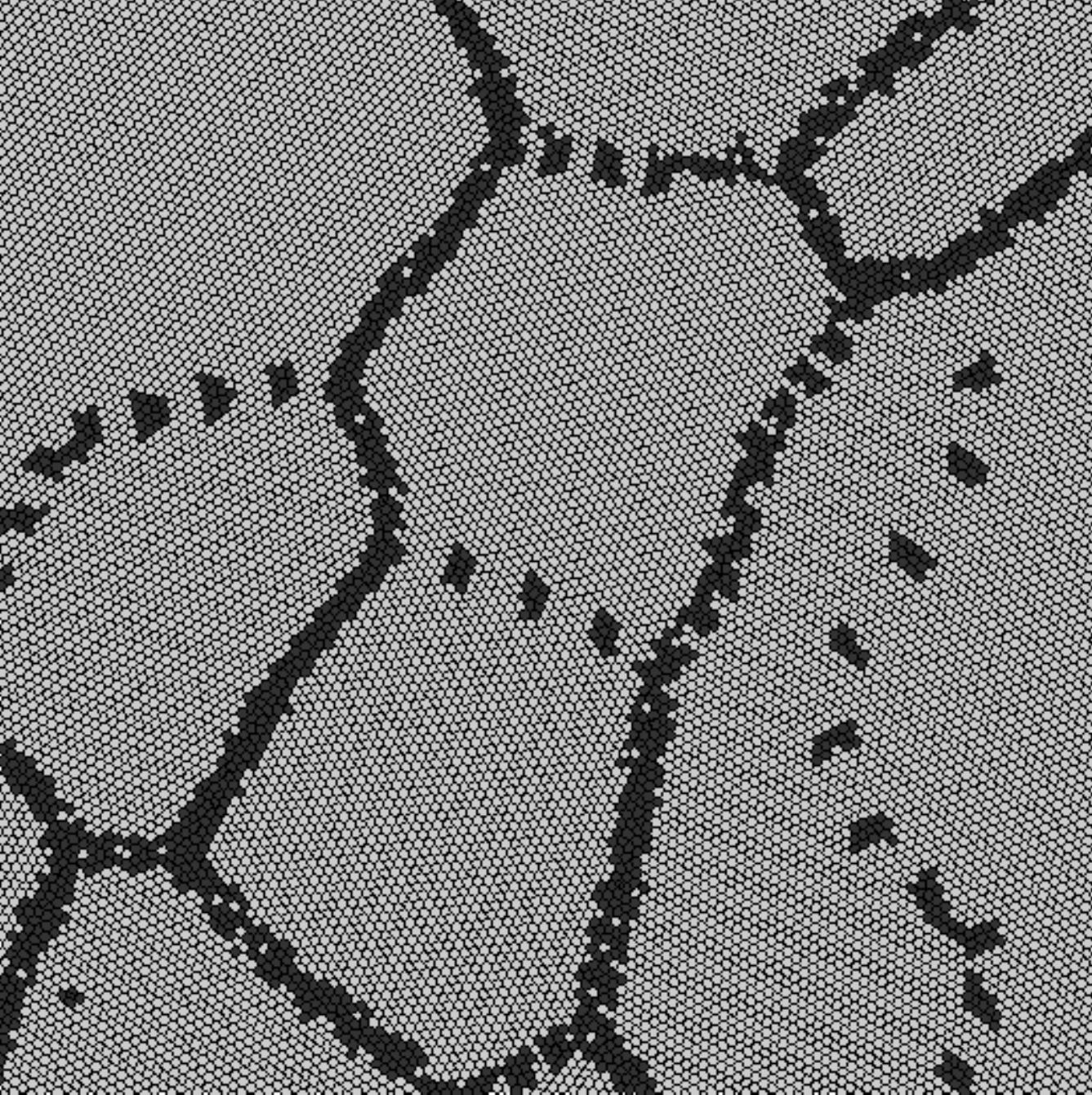}}
	
	\subfloat[]{\includegraphics[width=0.3\textwidth]{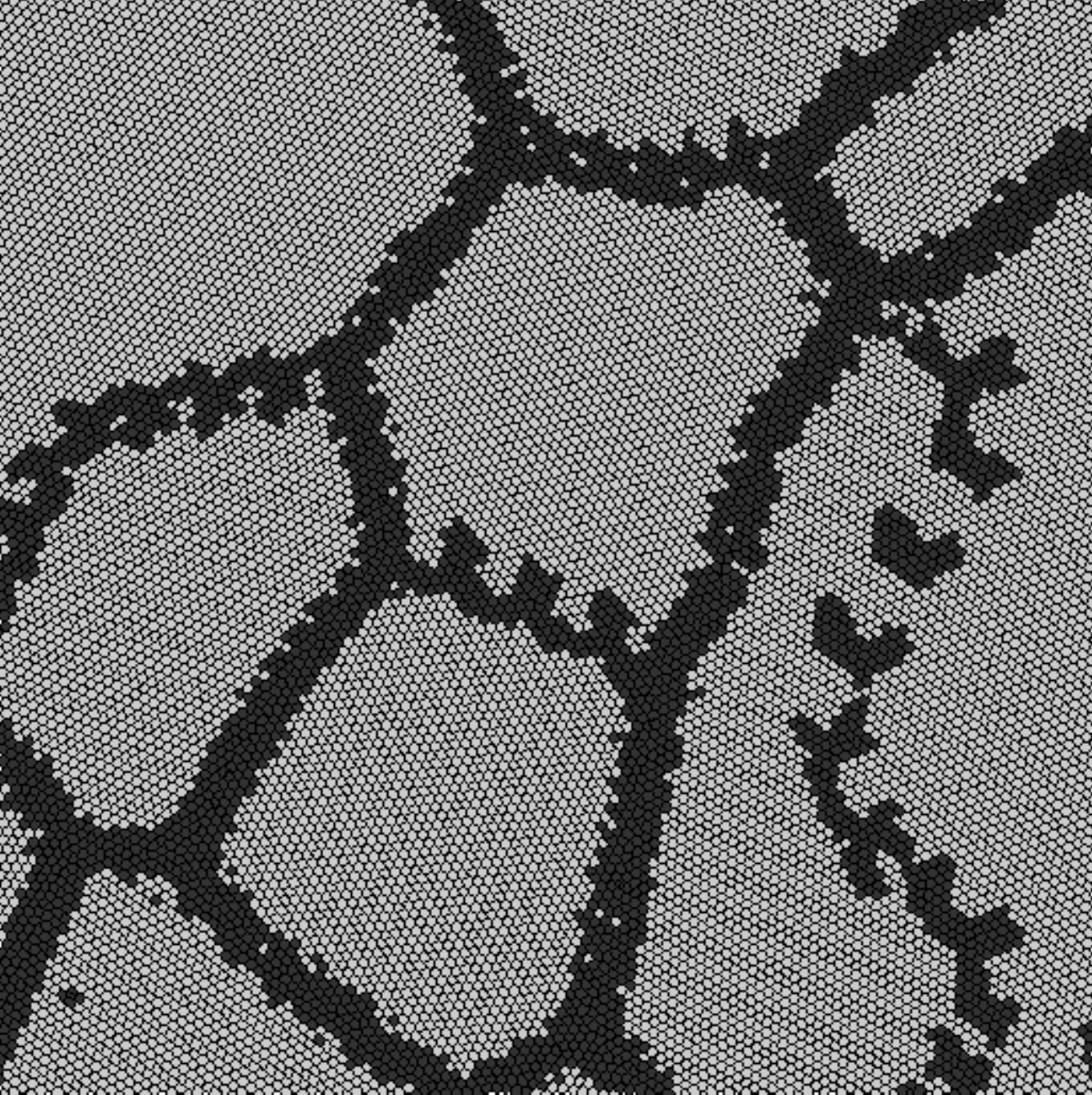}}\enskip
	\subfloat[]{\includegraphics[width=0.3\textwidth]{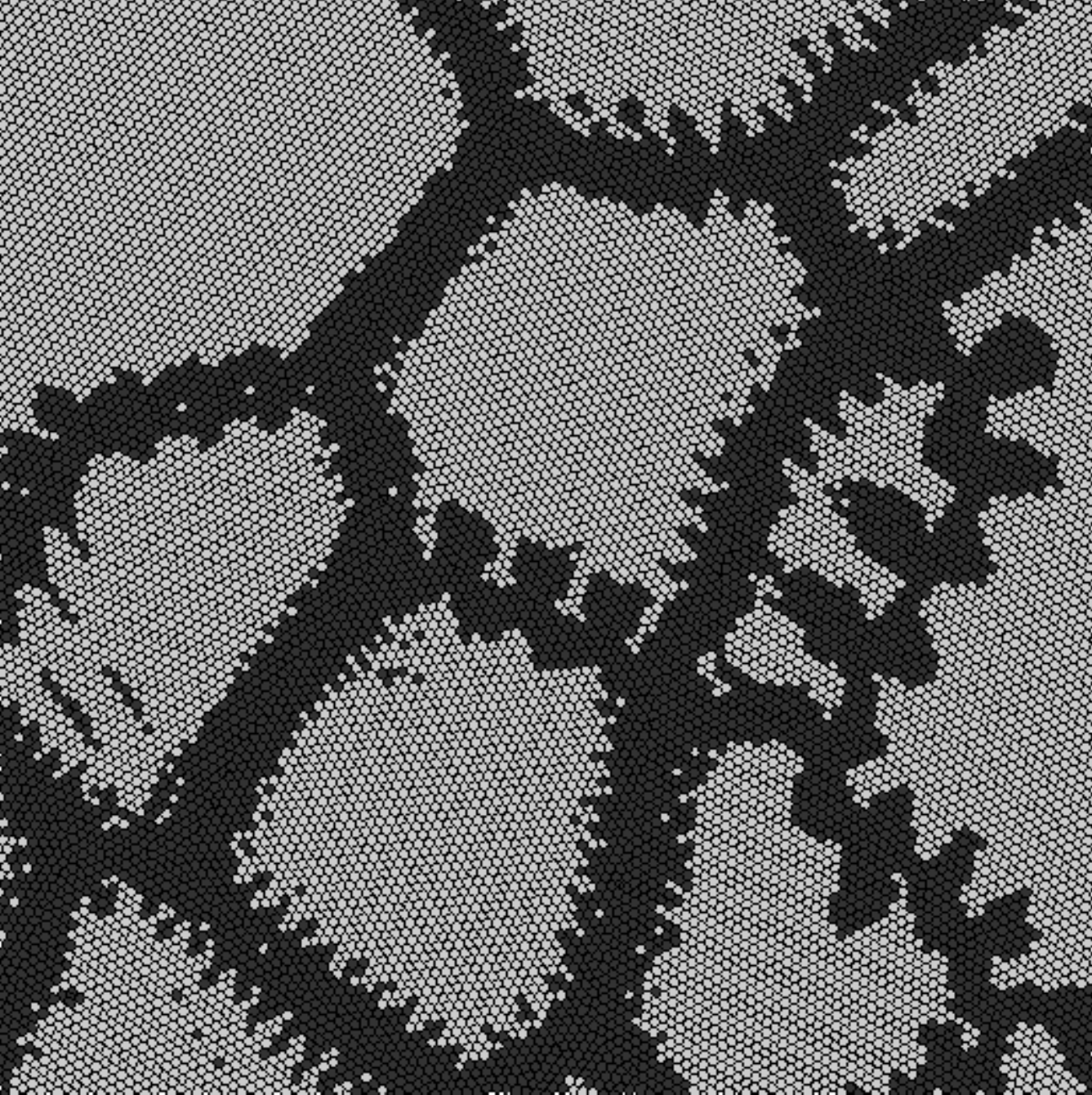}}
	\caption{Visualization of local lattice orientations in a PFC simulation (a). Grain boundary masks using $\gamma$ equal to $0.02, 0.005 \midand 0.002$ in (b,c,d) respectively.}
	\label{fig:QThreshold}
\end{figure}

A simple measure of closeness between the Voronoi cell and the regular hexagon is as follows. The area $A$ and perimeter $P$ of the Voronoi regions can be computed using methods such as the Shoelace formula \cite{BRADEN_Shoelace}. From these properties, one can compute the isoperimetric ratio $Q = \frac{4\pi A}{P^2}$ of the region. This parameter is positive and less than 1 by the isoperimetric inequality, that value being achieved only by a perfect circle. A similar result is that within the class of $N$-sided polygons, the regular $N$-gon achieves the maximum isoperimetric ratio $Q_N = \frac{\pi/N}{\tan(\pi/N)}$. This means that the cell of an atom is close to a regular hexagon if $Q$ is close to $Q_6$. To characterize defects and atoms close to grain boundaries, we will label any atom for which $Q_6 - Q$ is larger than some threshold $\gamma$ as ``bad''. In contrast, if an atom has six neighbors and its region is close enough to a perfect hexagon, it is called a ``good'' atom. The value of $\gamma$ then sets how different atoms close to boundaries are expected to behave compared to atoms deep in a perfect lattice. This is illustrated in figure \ref{fig:QThreshold}: notice how increasing $\gamma$ from (b) to (d) progressively thickens the boundaries and closes most gaps between grains until the boundaries become too thick and start invading the grains proper. The final results are not sensitive to $\gamma$ between these extremes. From this figure, the partition between good and bad atoms is clear: it gives a preliminary mask that can be used to ignore all misleading orientations near grain boundaries, improving the accuracy in the detected grain angle and boundaries.

Since this approach is similar to the Voronoi Analysis detailed in \cite{STUKOWSKI_OVITO}, it suffers from issues that reduce its applicability to other lattice types or to 3d. In such cases, as is done in \cite{PANZARINO_Tracking}, this step may be overhauled with similar methods such as Common Neighbor Analysis or Centrosymmetry Parameter.

\subsection{Flood Fill Procedure}
\label{sub:FloodFill}
Grains must now be identified by connecting neighboring good atoms based on their orientation. This step is similar to extracting connected components on a square grid, the main difference being that the algorithm works on the graph of atoms given by connecting neighboring atoms. These may be detected by connecting atoms that share a Voronoi vertex or, more efficiently, by connecting atoms that share a triangle in the Delaunay triangulation \cite{OKABE_VoronoiDelaunay} of the atomic positions.

A grain is found by picking a random good atom and probing its six neighbors. If a neighbor is good and its orientation is sufficiently close to that of the starting atom, it is accepted into the grain. How small this misorientation can be is set by the misorientation threshold $\theta$. The grain having grown by one shell of neighbors, a preliminary grain orientation is computed by averaging the orientations of accepted atoms. This process iterates for all new atoms with the difference that orientations are now compared to the preliminary grain orientation. When all neighbors of the grain are bad atoms or their orientation differences are larger than $\theta$, the number of atoms is computed. If it is smaller than an atom number threshold $\alpha$, the grain is deleted and its atoms are marked as bad. This ensures that spurious small clusters of good atoms, especially in highly stressed regions, are not falsely recognized as grains.

The procedure iterates until all good atoms have been associated to a grain or marked as bad, resulting in a partition of atoms in either grains or the thick grain boundary network similar to figure \ref{fig:QThreshold} (c). As long as no true grain has been missed, we can assume the ``sharp'' grain boundaries lies exactly in the middle of the region of bad atoms between preliminary grains. This is done approximately by connecting a bad atom to the grain that contains most of its neighbors, ignoring those connected only to bad atoms. This process repeats until all atoms belong to a grain. Since all grains acquire a few new atoms, the smallest grain detected will be slightly bigger than $\alpha$. An appropriate choice for this threshold is then any value smaller than half the size of the smallest feature that is represented in the input. Since very small clusters of good atoms may be present within the boundary network, $\alpha$ must still be larger than some minimum value to ensure that not too many false positives are detected.

\subsection{Measurement of Grain Properties}
\label{sub:Measurement}
Measuring grain properties turns out to be quite simple once grains are identified and their atoms known. Let us call boundary atoms those atoms that have a neighbor inside a different grain. We now describe how to compute several grain properties using only the Voronoi area of atoms.

\textbf{Grain area}: The area of a grain is the sum of the area of the Voronoi regions of its atoms. This property is usually expressed through the normalized reduced area, computed by first taking the root of the areas then dividing by the average.

\textbf{Grain perimeter}: The perimeter of a grain is the sum of the area of all boundary atoms that are connected to the grain, divided by a characteristic boundary thickness. This sum then includes all boundary atoms of the current grain plus the boundary atoms of grains that are in contact with it.

\begin{figure}[H]
	\centering
	\subfloat{\includegraphics[scale = 0.5]{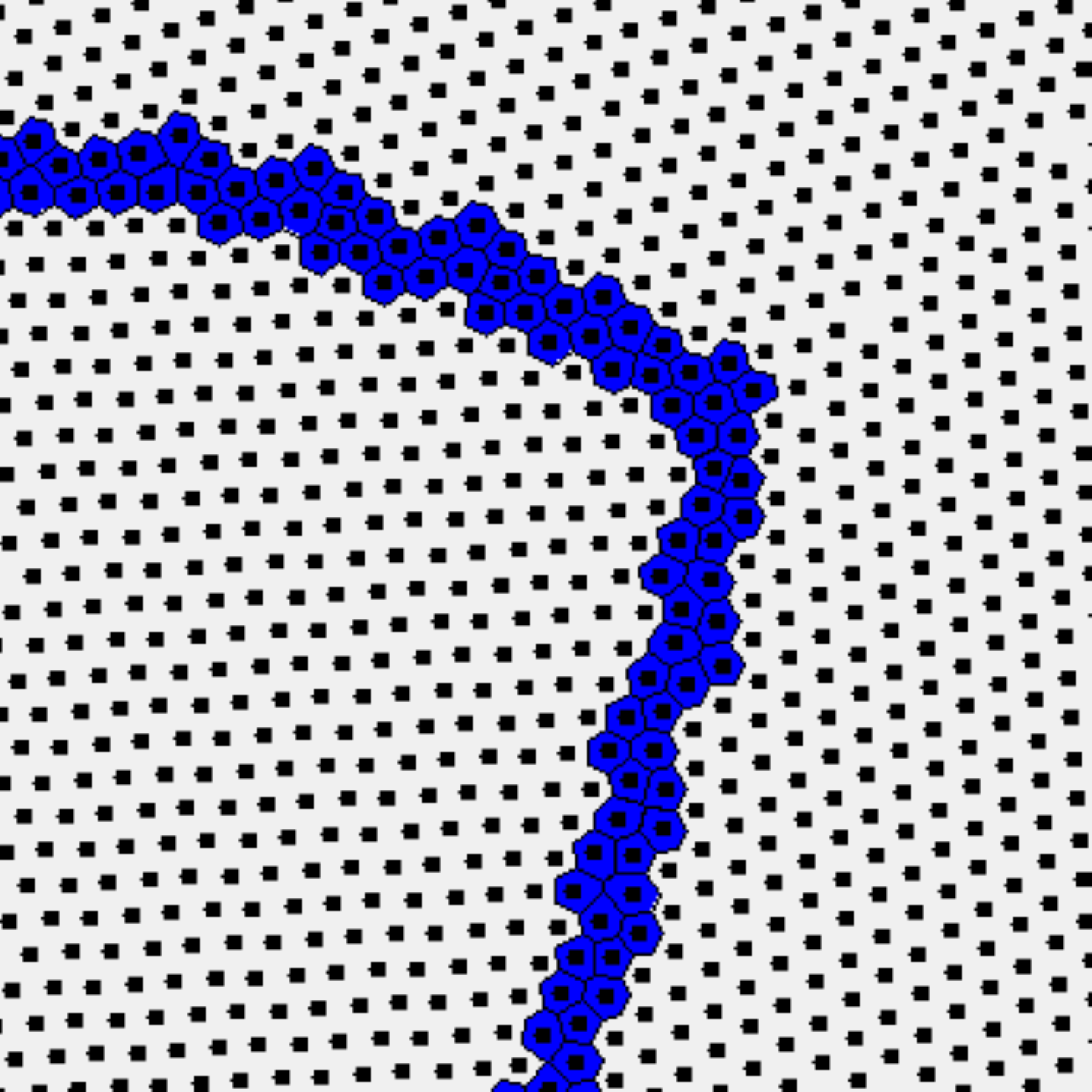}}
	\caption{Detail of a PFC grain boundary with extracted atomic positions in black and the Voronoi region of boundary region atoms colored in blue. }
	\label{fig:PerimBoundary}
\end{figure}

Several issues arise when attempting to compute grain perimeter since a boundary is an extended region ambiguously defined by discrete atomic positions. In addition, similarly to the coastline paradox, one could either trace a smooth line \textit{somewhere} through a grain boundary or one could trace the line connecting boundary atoms one by one. While the first approach is more sensible from a materials science point of view, only the second is unambiguous. Naturally, the perimeters recorded vary between the two approaches. Since there is no clear notion of perimeter, we use a geometrically unambiguous quantity to approximate the length of grain boundaries. Taking PFC as a concrete example, grain boundaries are very thin with a thickness comparable to the interatomic distance $d$. The region defined by the Voronoi regions of boundary atoms is then roughly two atoms thick, illustrated in figure \ref{fig:PerimBoundary}. Thus, dividing its area by $2d$ transforms the quantity into a one dimensional value akin to perimeter. While $d$ is not necessarily fixed in space or time, we simply assume that all atoms have a perfectly hexagonal Voronoi region and equally divide the domain so that given the domain area $A$ and the number of atoms $N_a$, one has that $d \approx \sqrt{2A/(\sqrt{3} N_a)}$ gives reasonable results as will be shown later.

Any other method of relating the area of the diffuse boundary to the perimeter could be used and accuracy might be gained by using a relationship that is better informed by the exact geometry of expected interfaces. An alternative and unambiguous characterization of area and perimeter in atomistic simulations may simply be the number of atoms inside a grain and at the boundary respectively. While this would not as well extend to mesoscopic experiments, it could be used to compare different atomistic simulations directly and unambiguously.

\textbf{Grain isoperimetric ratio}: This ratio is computed from the area and the perimeter. The closer this value is to 1, the closer a grain is to a circle. In particular, oscillations in the boundary quickly decrease this value. This provides a measure not only of grain circularity but also of the roughness of its interface. Our method does not provide a geometric isoperimetric ratio since area and perimeter do not represent the same curve, hence the ratio may be larger than 1.

\textbf{Grain coordination number}: The coordination number of a grain is its number of neighbors, computed by counting the number of grains with neighbor atoms in a given grain.

\textbf{Grain-grain area ratio}: Given the areas $A_1 \midand A_2$ of two neighbor grains, the area ratio is defined as the minimum between $A_1/A_2 \midand A_2/A_1$ to keep the ratio below 1. This property is an attempt to capture geometric correlations between neighbors. For example, if a grain has grown in time, its area must have been part of neighboring grains, giving rise to local imbalances in area. The evolution in time of the distribution of area ratios can then reveal whether this process ``balances out'' across the structure or whether large grains grow systematically to the detriment of very small neighbor grains.

\textbf{Grain-grain interface length, misorientation and GBCD}: As with the perimeter, the area of the Voronoi regions of \textit{only} the boundary atoms connecting two neighbor grains may be summed and divided by $2d$, measuring the length of the interface between the grains. The misorientation between two grains may be calculated using the periodic angular difference between the grain orientations. A derived property is then the GBCD \cite{HOLM_GBCD} which is computed by summing the interface lengths of grains whose misorientation falls in a given angular bin. In contrast to the bare misorientation distribution, this produces a histogram weighted by interface length rather than number which highlights preferred misorientations according to a geometrical metric.

\subsection{Post Processing Grains}
\label{sub:PostProcessing}
\begin{figure}[H]
	\centering
	\subfloat[]{\includegraphics[width=0.3\textwidth]{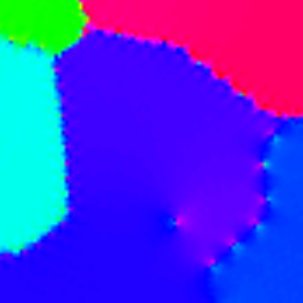}}\enskip
	\subfloat[]{\includegraphics[width=0.3\textwidth]{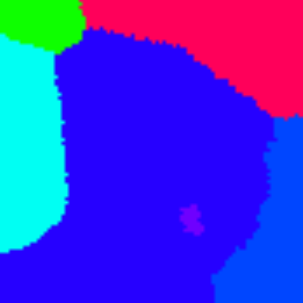}}
	
	\subfloat[]{\includegraphics[width=0.3\textwidth]{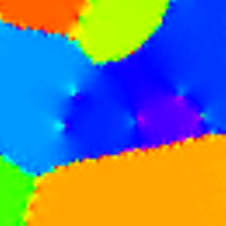}}\enskip
	\subfloat[]{\includegraphics[width=0.3\textwidth]{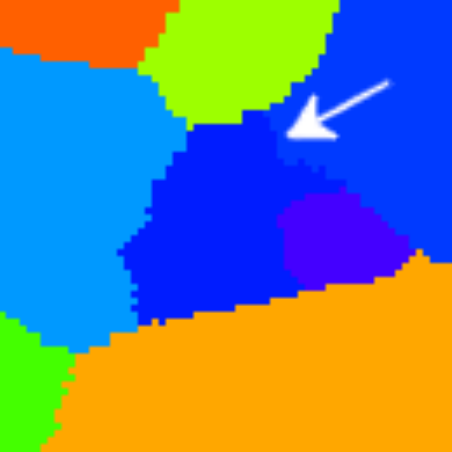}}
	\caption{Visualization of local lattice orientations in a PFC simulation (a, c). Corresponding visualizations of extracted grain orientations showing a single-neighbor grain (b) and a boundary with misorientation smaller than the threshold (d, white arrow).}
	\label{fig:GrainExtractionDefect}
\end{figure}

Some issues in the original input image may lead to the detection of spurious grains. This is a problem when using very small thresholds in the hope of recognizing all true grains. The principal cause of concern is the presence of defects in ambiguous low misorientation boundaries as shown in figure \ref{fig:GrainExtractionDefect}. In (a), it is quite ambiguous whether there are two purple grains or a single one. Upon closer inspection, the misorientation between the dark and light purple is $2.6^\circ$ which is slightly above the chosen threshold $\theta = 2.5^\circ$. However, there is a very smooth transition between the two orientations suggesting that this is a single grain under stress due to the large defect, hence, it should not be split. While the algorithm correctly identifies the whole smooth region as a single grain, the defect influences a large set of good atoms. Such regions around point defects are sometimes counted as spurious grains, especially in large grains, as in (b). These artefacts may be identified by detecting grains that have only one neighbor, which should be extremely rare in usual circumstances.

A similar issue sometimes arises at the boundary between two grains. Unlike defects, the problem is that part of the whole diffuse boundary is detected as an independent grain. Such cases cannot be detected on the basis of the coordination number but they usually have a very small misorientation to one of their neighbor. An example of such an artefact is shown in figure \ref{fig:GrainExtractionDefect} (c). While the diffuse region contains a dark blue and a purple grain, the misorientation between the blue half and the dark blue grain to its right is of only $1^\circ$. The boundary marked by the white arrow in (d) is then spurious and must be deleted. Such cases are detected easily by finding all grain to grain misorientations smaller than $\theta$ which then becomes an exact threshold. In both cases presented above, the spurious grain is simply deleted and its atoms are reassigned to the relevant neighbor.

\section{Validation}
In this section, we validate the results of our atomic grain extraction procedure. The first issue is to verify that we properly account for almost all grains present in the input and identify as few spurious grains as possible. The definition of a grain is obviously subject to interpretation hence our goal is to ensure that the numerical scheme essentially reproduces a human manual segmentation. The second issue is that once all grains are properly identified, their geometric properties must be faithful to the input. This is simple to verify for most properties except those based on interface lengths because such a definition is ambiguous as described before. Our source of validation is two-fold: first, we use results from PFC simulations, which we briefly summarize, to have a ``natural'' grain distribution that can be hand segmented. Second, we produce artificial grain distributions in which grains are known geometric shapes whose number and properties are unambiguous. In both validation sources, there are targets for the grain structure and geometric properties that the numerical scheme must recover. For completeness, we also compare the atom and grid based approaches to show that the proposed method is more robust. To make the process simpler, we summarize the four numerical parameters used in the scheme:
\begin{itemize}
	\item Atom extraction level $h$: Level at which to trace atomic contours in the input image.
	\item Threshold isoperimetric ratio $\gamma$: How close to a perfect hexagon the Voronoi cell of atom should be.
	\item Threshold misorientation $\theta$: The minimum grain misorientation that the scheme can accurately identify.
	\item Threshold number of atoms $\alpha$: The minimum number of good atoms for a grain to be accepted initially.
\end{itemize}
When testing these parameters, we shall focus our attention on $\theta \midand \alpha$ since they are the ones that influence the results significantly. We shall use cumulative density functions (CDF) to compare between data sets since there is no guarantee of a one-to-one map between recognized and actual grains. Unlike a histogram, a CDF is unique and can be used even for very small data sets.

\subsection{PFC Evolution}
While many more elaborate PFC-like models have been developed, ranging from the prototypical evolution of \cite{ELDER_Elasticity} to those modelling dendritic solidification \cite{OFORI_Dendrites} and even graphene \cite{SEYMOUR_Graphene}, its simplest 2d form gives the evolution of a phase field $u$ according to the partial differential equation 
\begin{equation}
u_t = \lapl \left( (\lapl + q_0^2)^2 u + u^3 - \beta u \right) \ .
\end{equation}
Such an evolution conserves the average phase $\langle u \rangle = m$. The parameter $\beta$ can be thought of as an inverse temperature while $q_0$ sets an atomic length scale. The phase diagram of the system in $(m, \beta)$ can be divided into three regions: a ``liquid'' state where $u = m$, a ``rolls'' state $u = m + A \sin(q x)$ and finally a state where $u$ is a superposition of three rolls misoriented by $60^\circ$ forming a hexagonal lattice. In this regime, an initially noisy phase field quickly evolves to form local ``bumps'' representing atoms. These arrange locally in small hexagonal clusters with interatomic distance $d(q_0) = 4\pi/(\sqrt{3} q_0)$. The subsequent evolution is at a macroscopic level as clusters become grains and coarsen. This PFC system gives a prototypical evolution with a single crystalline lattice type and is thus the ideal test for our method.

To simulate the evolution, we use the regularized semi-spectral numerical scheme developed in \cite{ELSEY_Scheme}. We fix a square domain with periodic boundary conditions and choose the same PFC parameters $(m, \beta) = (0.07, 0.025)$ along with the regularization parameter $C = 2\beta = 0.05$ and $\tau = 1000$. Fixing $q_0 = 1$, the numerical domain size is given by $L = 1024 d(1) \approx 7429$ with $8192$ grid points. Defined this way, the domain supports roughly $1024$ atoms in a dimension aligned with the lattice and each atom is resolved by approximately $8^2$ pixels. For the PFC parameters above, we chose an atom extraction level of $-0.035$ and an isoperimetric ratio threshold of $0.0005$.

\subsection{Comparison With PFC Grain Distributions}
\begin{figure}[H]
	\centering
	\subfloat[]{\includegraphics[width=0.4\textwidth]{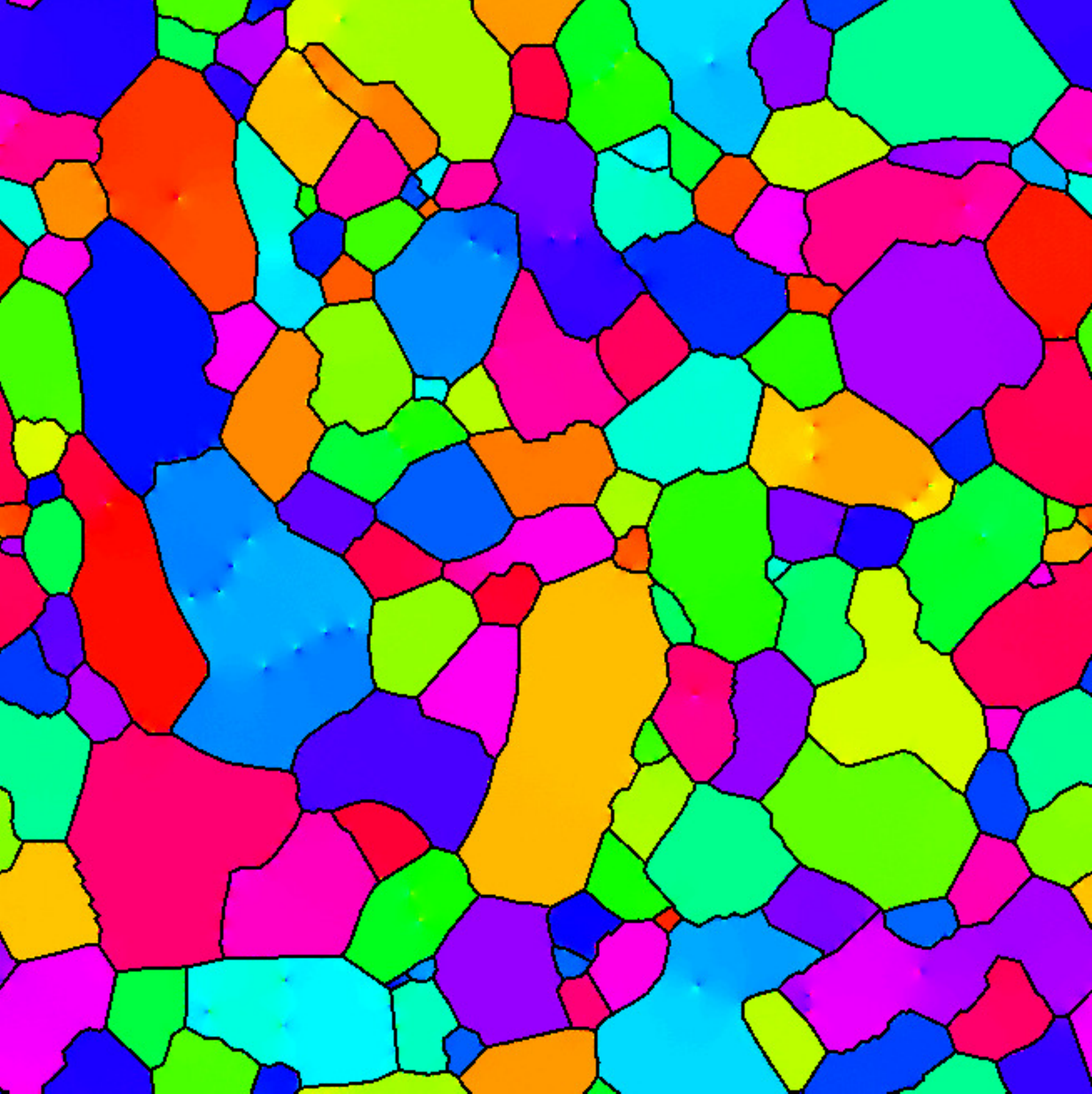}}\enskip
	\subfloat[]{\includegraphics[width=0.4\textwidth]{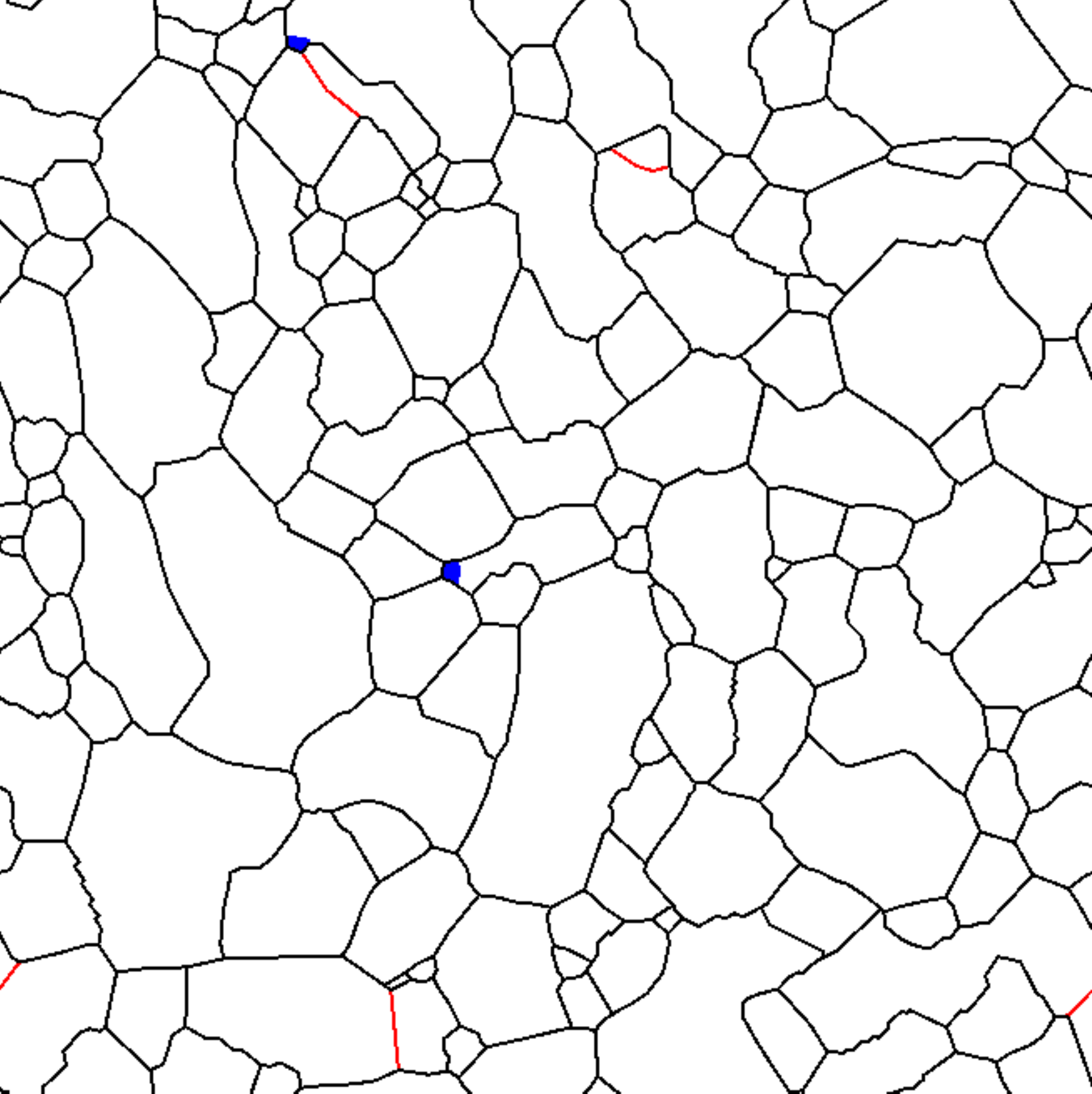}}
	\caption{Visualization of local lattice orientations in the late distribution with manually segmented grain boundary network in black (a). Comparison between the manual and numerical segmentations (b); common boundaries are drawn in black, two spurious grains are drawn in blue and four undetected boundaries are drawn in red.}
	\label{fig:GrainBoundariesLate}
\end{figure}

Our first comparison consists in the evolution of a noisy phase field according to the scheme above. The phase was saved after $228 \midand 40000$ time steps, corresponding to roughly $1.25$ million atoms divided in $1800 \midand 130$ grains respectively. For clarity, we shall call these the early and late snapshots. To manually segment the distributions, atomic orientations were projected on an image over which grain boundaries were carefully traced as polygons without overfitting atomic structural details. Since the early distribution contains so many grains, only $459$ grains were extracted which represents a little more than a quarter of the entire domain. The target number of grains should therefore be close to $1830$ with an average perimeter of $680$. Note that average area is redundant since it equals the area of the domain divided by the number of grains. Because low misorientation boundaries are ambiguous, especially in large grains, it is not realistic to trace \textit{all} boundary that are visible or implied by a line of defect; one should ideally consider the misorientation between the two grains as well as the openness of the line of defects. Without accepting such poorly defined porous boundaries, the manual boundary network in the late distribution is shown in figure \ref{fig:GrainBoundariesLate} (a). According to this segmentation, the late distribution contains $129$ grains while the average perimeter is $2426$. The only regions prone to be ambiguous are those in which misorientations are smaller than about $5^\circ$ as otherwise grain boundaries are very sharp and clear.

We processed the same PFC phase fields using our numerical scheme with a variety of parameter sets. Since the hand segmentation process becomes quite ambiguous below roughly $5^\circ$ misorientations, we chose $\theta = 2.5^\circ$. Choosing the area threshold requires more thought because the same parameters should ideally be used irregardless of feature size. The smallest features found in the early and late distributions contained $75 \midand 200$ atoms respectively, suggesting a choice of $40$ for $\alpha$. Since this choice must be valid in both cases, we must allow for some omissions in the early distribution and some false positives late in time. With these choices, the numerical scheme detects $1739$ grains with an average perimeter of $678.6$ in the early distribution and $127$ grains with an average perimeter of $2409$ in the late distribution.

\begin{figure}[H]
	\centering
	\subfloat[]{\includegraphics[width=0.4\textwidth]{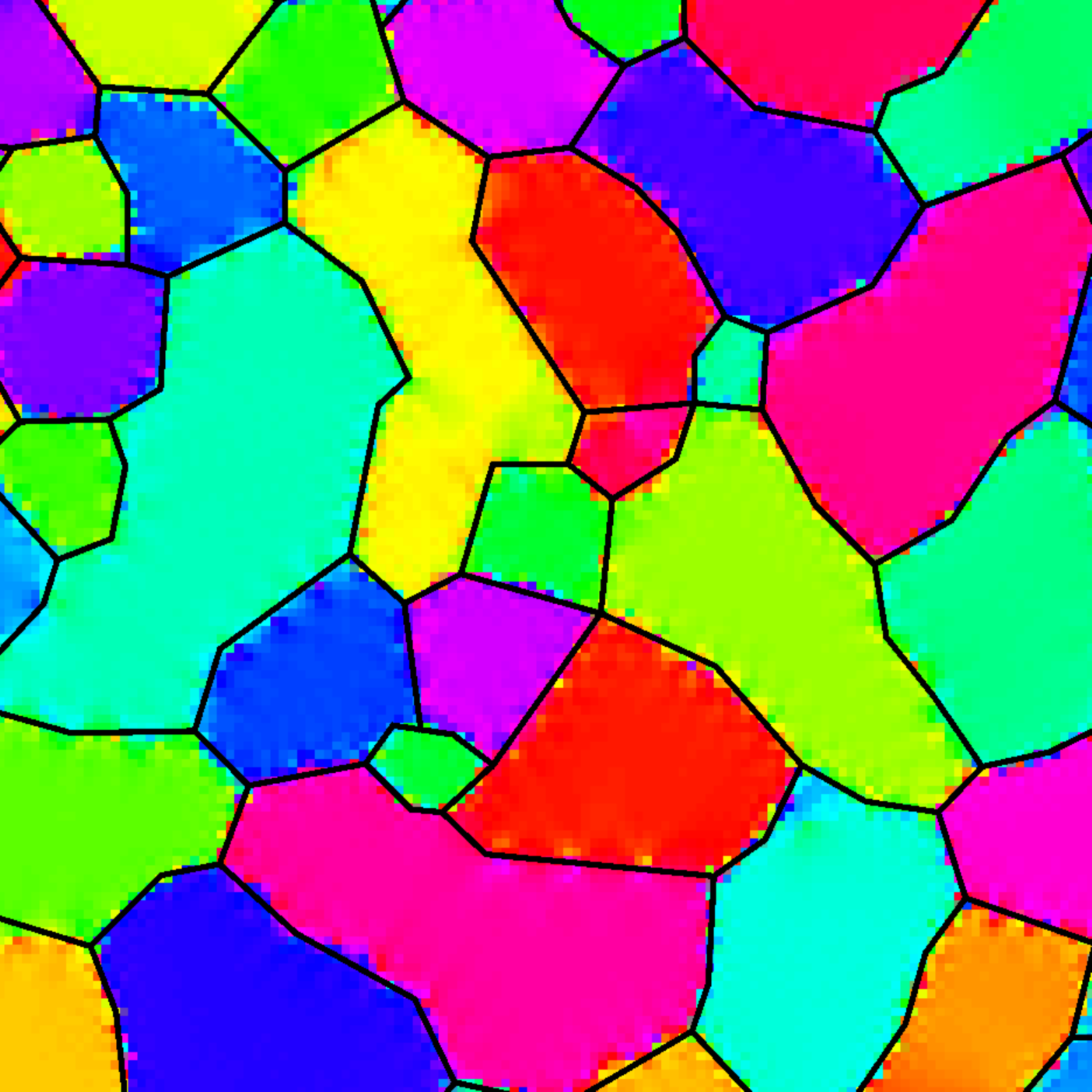}}\enskip
	\subfloat[]{\includegraphics[width=0.4\textwidth]{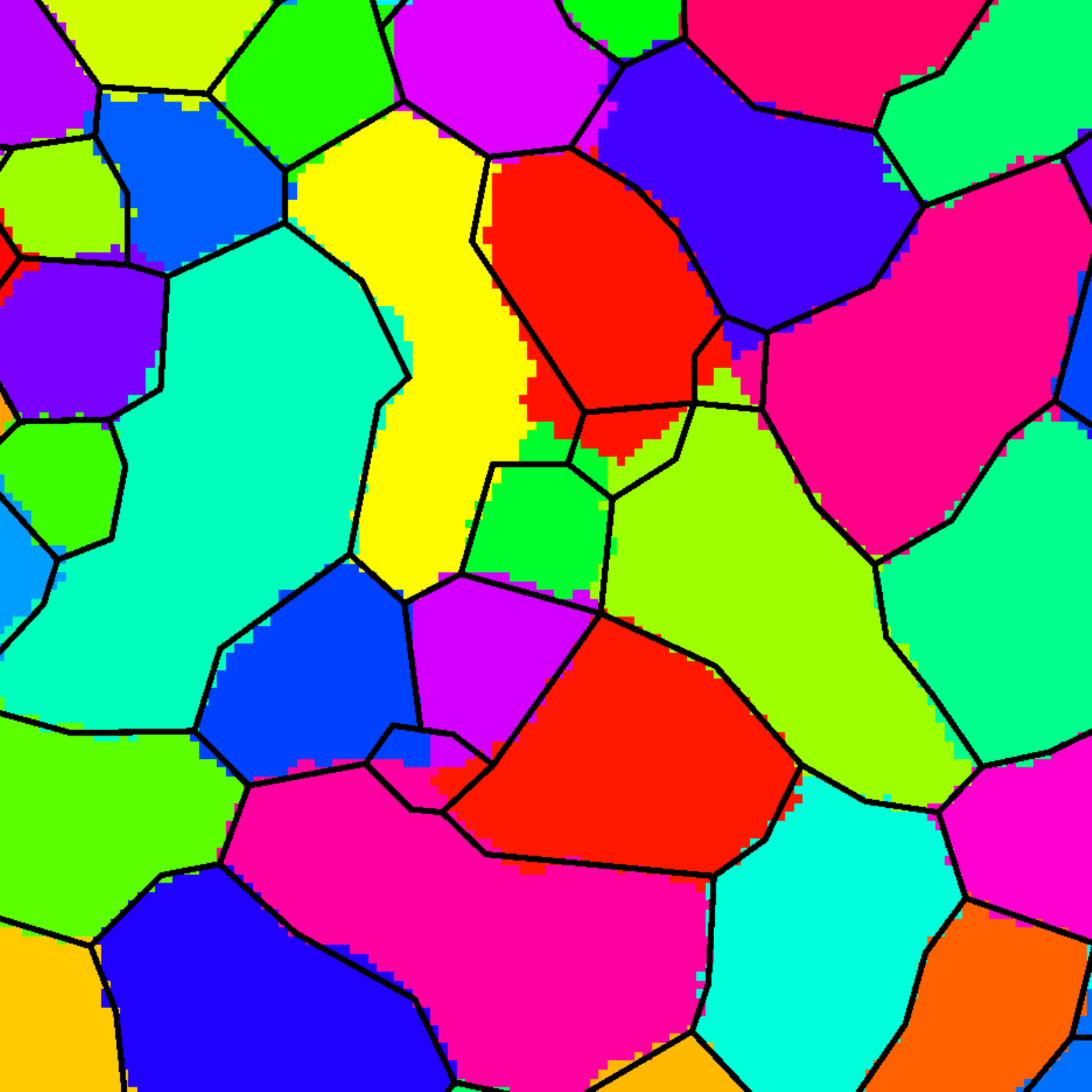}}
	\caption{Detail of the visualization of local lattice orientations in the late distribution with manually segmented grain boundary network in black (a). Comparison between the manual and numerical segmentations (b); the numerically extracted grains are visualized in color while the manual segmentation is drawn in black.}
	\label{fig:GrainBoundariesEarly}
\end{figure}

Let us first compare the boundary networks in the late distribution, shown in figure \ref{fig:GrainBoundariesLate} (b). In total, six boundaries are in disagreement, four being found only in the hand segmentation and two spurious grains being found numerically. These boundaries are all ambiguous and porous with a misorientation lying between $2 \midand 3$ degrees. The scheme is then able to detect almost all hand segmented grains while identifying a minimal number of artefacts. A similar comparison for the early distribution is done in figure \ref{fig:GrainBoundariesEarly}. The left image shows a small portion of the local atomic orientation with the manual segmentation while the second compares the numerical and manual segmentations. The agreement is quite good considering the small working scale. Three very small grains have been missed by the numerical algorithm since their size is smaller than the threshold $\alpha$. These could have been detected with a smaller threshold at the cost of introducing spurious grains. We note that taking the difference between the raw orientation image and the grain orientation image, one can easily see regions where angles are poorly captured. In particular, missed grains appear as particularly bright regions. Counting these gives approximately $90$ missed grains in total which is consistent with the difference between the expected and recovered number of grains. This means that roughly $5\%$ of grains were too small to be detected by the method.

\begin{figure}[H]
	\centering
	\subfloat[]{\includegraphics[width=0.45\textwidth]{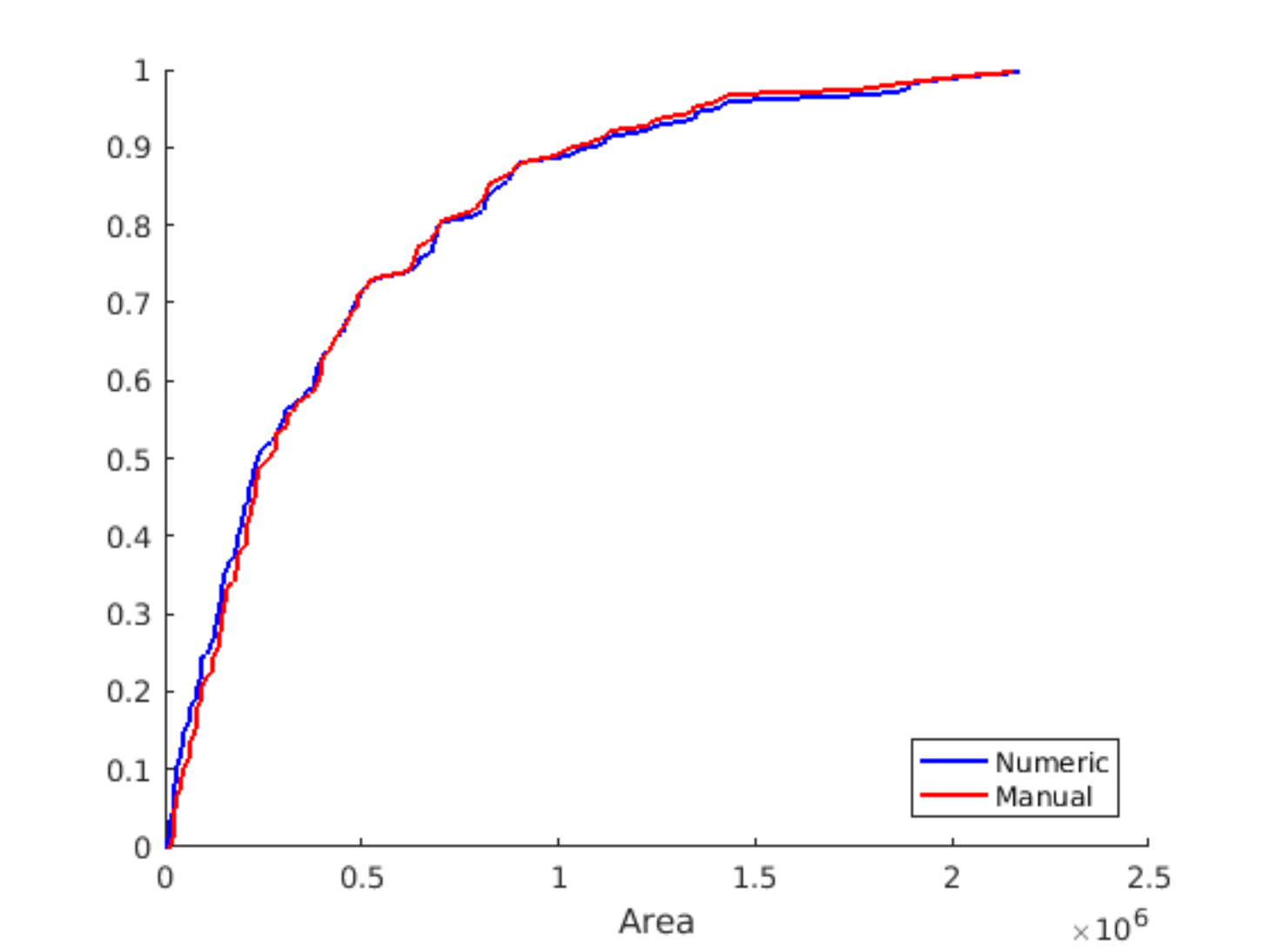}}\enskip
	\subfloat[]{\includegraphics[width=0.45\textwidth]{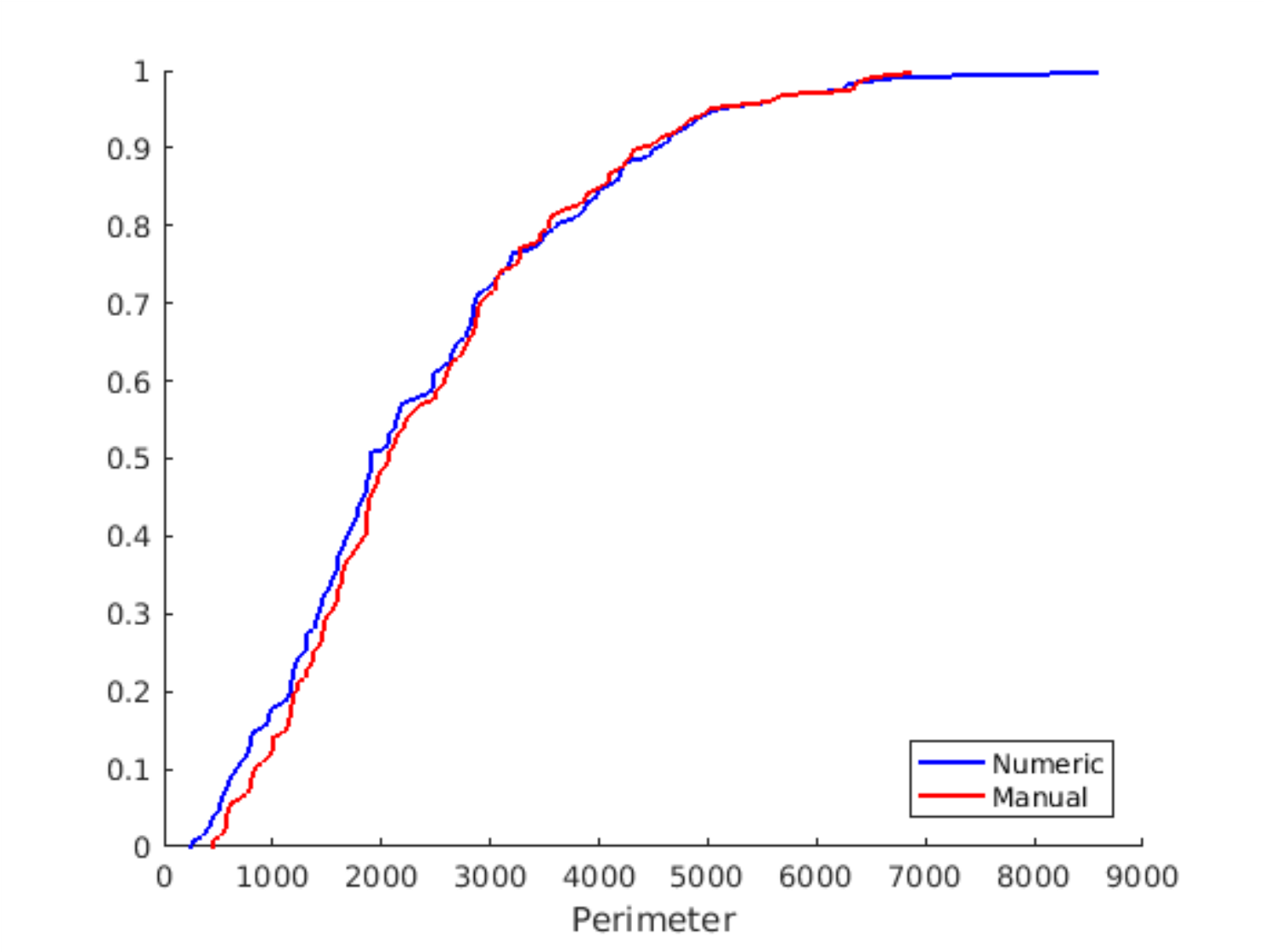}}\enskip
	\subfloat[]{\includegraphics[width=0.45\textwidth]{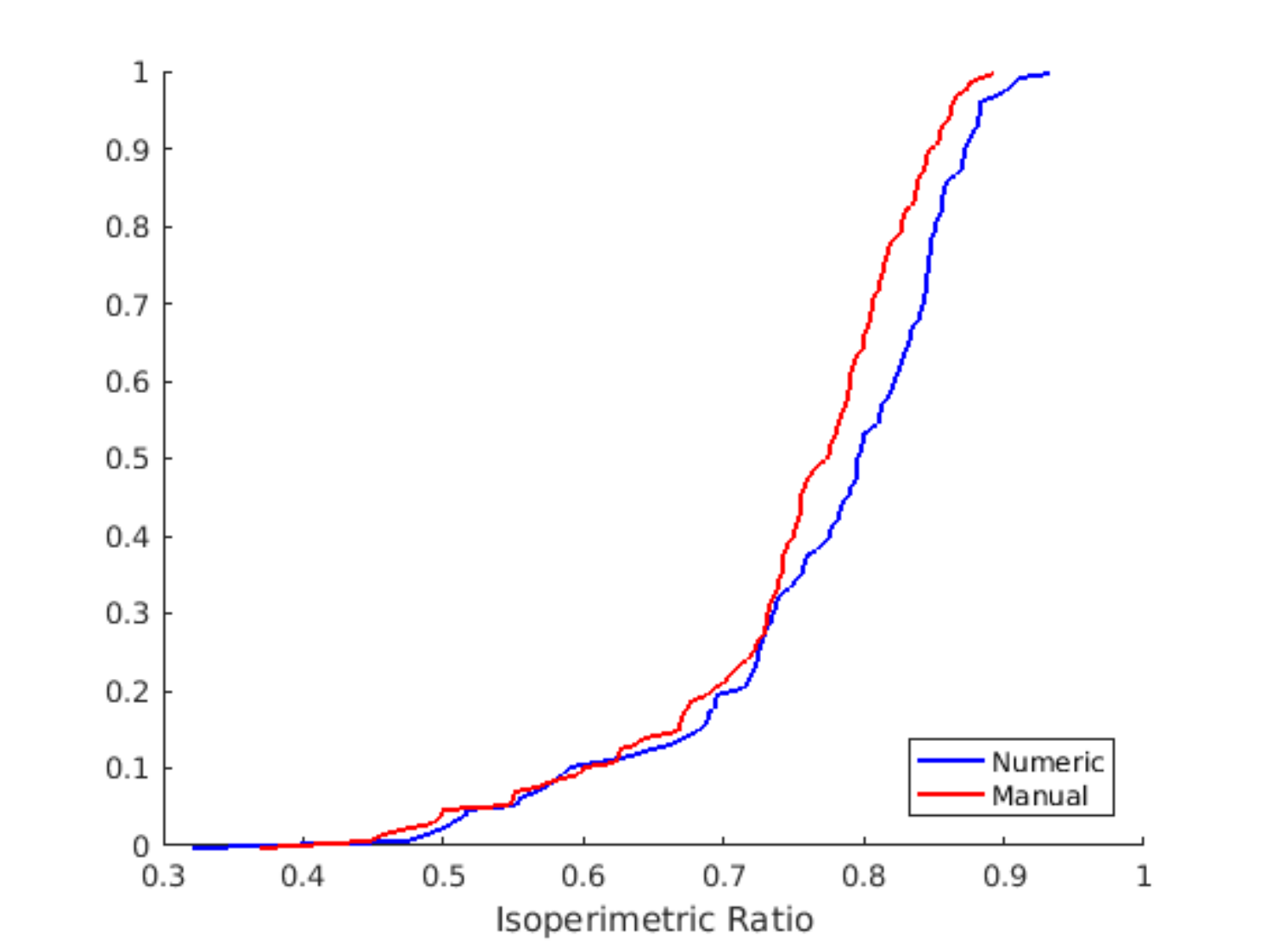}}\enskip
	\caption{Comparison between the area (a), perimeter (b) and isoperimetric ratio (c) CDFs of the hand segmentation in red with the numerical extraction procedure in blue. Note that in (b), the grain with the largest numerical perimeter corresponds to the magenta/purple bottom right grain that was not properly split in figure \ref{fig:GrainBoundariesLate}.}
	\label{fig:AgreementAreaPerimIPR}
\end{figure}

It remains to compare the geometric properties. We focus on comparing the area, perimeter and isoperimetric ratio since other properties mostly depend on the networks being well recovered. The agreement between the CDFs of the manual and numerical segmentations is shown in figure \ref{fig:AgreementAreaPerimIPR} for the late distribution. While the area CDFs agree remarkably well, there are some differences in comparing the perimeter and isoperimetric ratios that can be traced to very small grains. The perimeter of these grains is generally underestimated, thus causing the shift in the perimeter comparison below $2000$ and that in the isoperimetric ratio above $0.75$. If one matches the numerically detected grains to their counterpart in the hand segmentation and ignores outliers, for example when two grains have been merged, one finds that area is captured with less than $1\%$ error while the perimeters are on average underestimated by about $2\%$. This is especially the case for very small grains whose perimeter is sometimes underestimated by as much as $20\%$. The comparison in the early distribution is very similar albeit with slightly poorer accuracy since grains are in general smaller. This highlights a slight underestimation of perimeters that depends on grain size.

While the chosen thresholds give rise to good agreement, it remains to determine how sensitive the results are to changing the parameters. Using the late distribution, we varied the thresholds and repeated the grain extraction procedure highlighting a simple characterization. As $\theta$ is increased, less grains are detected as expected. An increase of $0.5^\circ$ results in a reduction by approximately $6$ grains. However, it should be noted that below $2.5^\circ$, a large amount of spurious grains is detected suggesting that $2.5^\circ$ is the smallest threshold for which low misorientation boundaries can be characterized accurately.

The area threshold itself only affects the results when it is close in size to the smallest features present in the image and when it is small enough that many obvious artefacts are identified as grains. In the early distribution, the number of grains decreases steadily when $\alpha$ is increased between $30 \midand 70$. Comparing the boundary networks, it is clear that above $30$, the grains that fail to be detected are small but are not necessarily ambiguous. On the other hand, below $30$, the grains that are introduced in the segmentation are mostly artefacts within grain boundaries. The smallest threshold that does not introduce too many false positives is $30$, counting $1780$ grains. In the large distribution, several artefacts are recognized when $\alpha$ is less than $40$, otherwise the results are stable since the smallest features would start to be lost at a threshold above $100$. When looking at the boundary network using $30$ as the threshold, it is found that $5$ more spurious grains are found compared to the comparison shown above. This leads us to conclude that while an area threshold of $30$ would be ideal early in time, a threshold of $50$ is preferable later in time. To balance the method between these two quantities, we prefer to use $\alpha = 40$ even if this threshold does miss several grains early in time and risks introducing a few artefacts late in time. However, the CDFs obtained with any $\alpha$ between $30 \midand 50$ are mostly equivalent.

Finally, we note that since the flood-fill approach uses a random atom as its starting point, the results may differ each time the algorithm is run. Such differences are negligible, for example, upon 10 different runs of the algorithm on the late distribution, the number of grains found varied by at most 3 grains while the average perimeter varied between by less than $1\%$. The boundary networks and the CDF were similar in all cases and similar conclusions apply to the early distribution.

\subsection{Comparison With Voronoi Grain Distributions}
We now present a similar validation using an artificially constructed grain distribution so that areas and perimeters may be computed exactly. Given a target number $N$ of grains on a domain of a given size, we compute the Voronoi tessellation of $N$ randomly distributed points. Each Voronoi region is made to correspond to a grain whose angle is chosen randomly. Lattices at the appropriate orientation are placed onto the regions and passed through the PFC scheme to smooth out atoms at boundaries. Results are then extracted using the algorithm. To compare with the scales used in PFC simulations, we let $N$ vary between $125 \midand 2000$. Since grain misorientations below a certain threshold are ambiguous, we will not be able to recover all grains produced by the purely random Voronoi diagram. Instead, we can compute the proportion of grains in the diagram whose minimum misorientation is greater than the $2.5^\circ$ threshold and take this number of grains as the target. The fraction of such grains is found to be independent of grain size and equals about $40\%$. Thus, roughly half of these grains will be merged to a neighbor, giving a target grain number of $0.8 N$. Another approach, useful to compare the area and perimeter distributions directly, is to preemptively filter out grains with a small misorientation in the diagram and change their angle until they pose no such problem.

\begin{figure}[H]
	\centering
	\subfloat{\includegraphics[width=0.5\textwidth]{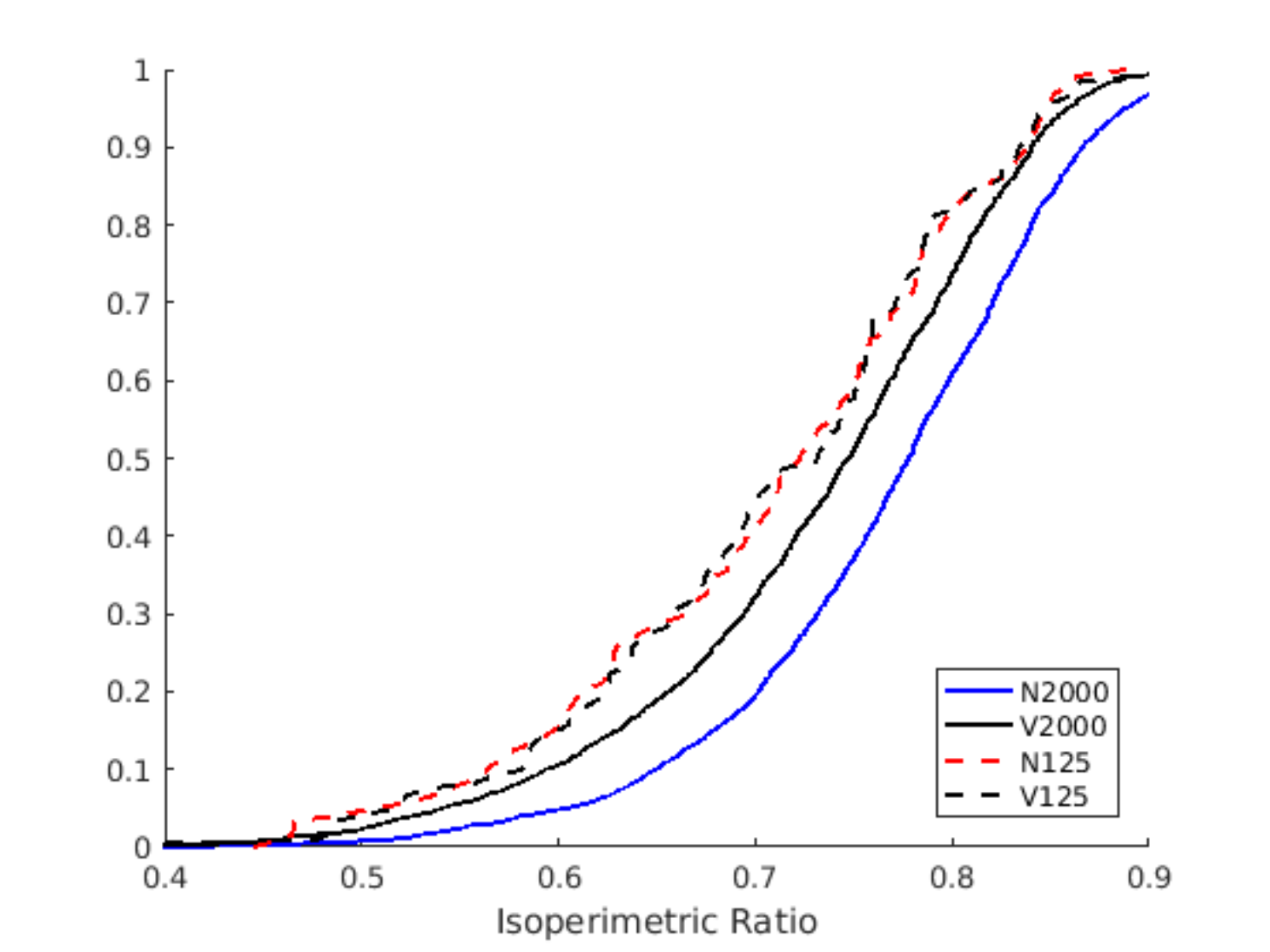}}
	\caption{Comparison between the isoperimetric ratio CDFs of the $2000 \midand 125$ grains Voronoi distributions in black and the numerically extracted results in blue and red. The dashed lines correspond to the $125$ grains distribution.}
	\label{fig:Voronoi_CDF}
\end{figure}

The algorithm is capable of finding the correct number of grains in both cases using the thresholds determined before. When the angles are not filtered, an average of $79\%$ grains are found across all lengthscales. This agrees with the expectation that a fifth of all grains cannot be resolved given this threshold. Obviously, the average perimeter is overestimated at roughly $115\%$. When angles are pre-filtered, an average of $99\%$ grains are detected at all lengthscales while the area and perimeter CDFs match closely. Random Voronoi diagrams rarely generate very small grains, unlike in the early PFC distribution presented before, so that almost all Voronoi grains may be detected at the chosen $N$. The average perimeter is recovered at $101\%$ for $N = 125$ and at $98\%$ for $N = 2000$. When matching grains, one finds that the error in the area is negligible and independent of grain size. On the contrary, the error in the perimeter depends on grain size and is consistent with the previous results: for $N = 125$, perimeter is recovered almost exactly while for $N = 2000$, grains are underestimated by $4\%$. Thus, the CDF of the isoperimetric ratios match almost exactly when there are $125$ grains but are spaced by $0.03$ when there are $2000$ grains as shown in figure \ref{fig:Voronoi_CDF}. These are quite slight deviations but when comparing the isoperimetric ratios on the same plot, this discrepancy does lead to confusion as the bias may be confused with evolution, meaning that isoperimetric ratio should at this point only be compared qualitatively.

\subsection{Comparison With Simple Geometric Shapes}
We now probe the accuracy of the scheme on very simple geometric shapes. In what follows, a grain whose boundary is a simple shape is created by placing a lattice at a given angle within another lattice at a different angle and passing this distribution through the PFC scheme to smooth out atoms. The grain extraction algorithm is then used to compute the geometry of the inner grain which is to be compared to known values. This was done for various various sizes, shapes and misorientations but the later did not affect the results. Overall, the area was always recovered within $\pm 0.5\%$ accuracy while the perimeter was recovered with similar accuracy for curved shapes and underestimated with a slight bias consistent with the results found above. The reason the curved shapes are precisely matched in perimeter is because they can only be represented as polygons using atoms. As such, the perimeter of the shape of the polygonal grain will be slightly larger than that of the true curved shape, cancelling out the inherent bias of the scheme.

\subsection{Comparison Between Atom and Grid Based Approaches}
Finally, we compare the atom and grid based approaches to highlight the advantages of the proposed method. We reuse the same phase fields obtained in the PFC and Voronoi comparisons above to compare the target number of grains and CDFs with both extraction methods. The grid based approach is implemented as described in section \ref{sub:GrainExtractionProblem}. Atomic orientations are computed as before and projected onto a grid of size $1024^2$. With this choice, pixels contain one or two atoms in a perfect lattice and $0$ or more at grain boundaries. If no atom is contained in a given pixel, it is assigned to the boundary mask by default instead of being needlessly interpolated. Otherwise, the boundary mask is computed using the quantity $4 |\nabla v|^2$ thresholded at the value $m = 0.33$. Similarly to the atom based approach, small grains are discarded when they contain less than $15$ pixels. These parameters are not as transparent as in the atom based approach and have simply been chosen to maximize the agreement over all comparisons presented below. Again, single-neighbor grains and those with neighbors with a neighbor misorientation smaller than $2.5^\circ$ have been removed.

When the early and late PFC snapshots are measured, $1696$ and $116$ grains are detected respectively, $2 \midand 8\%$ fewer than found using the atom based method. The missing grains are usually those with ambiguous misorientations close to the $2.5^\circ$ threshold. This does not change when the number of pixels threshold is decreased. On the other hand, $m$ may be chosen to improve the detection accuracy in the early or late snapshots at the expensive of dramatically decreasing the accuracy in the other. Even when $m$ is adapted to either snapshot, the number of detected grains is still smaller than using the atom based approach. This suggests that the thresholding of a simple measure of orientation inhomogeneities is insufficient to characterize grain boundaries especially near ambiguous grain boundaries. The grid based method performs much better on artificial Voronoi data, recovering as many grains as the atom based method for a large range of $m$. This is not surprising since in this artificial case, the orientation changes over a length of at most $2$ pixels by construction.

The extracted area and perimeter can also been compared to reveal that in general, areas are captured as accurately using both methods but the accuracy of the grid based method for the perimeter depends strongly on their shape. Since the perimeter computation traces a boundary in a pixelated image, sharp corners must be assumed to be projections of curved shapes. For this reason, while the perimeter of PFC grains or circular test shapes is slightly underestimated, that of very sharp Voronoi grains or square test shapes is severely underestimated. This effect is similarly exacerbated with smaller grains hence the grid based method does not in general capture perimeter as well as our calculation based on the area of boundary regions.

\section{Phase Field Crystal Sample Results}
\begin{figure}[H]
	\centering
	\subfloat[]{\includegraphics[width=0.45\textwidth]{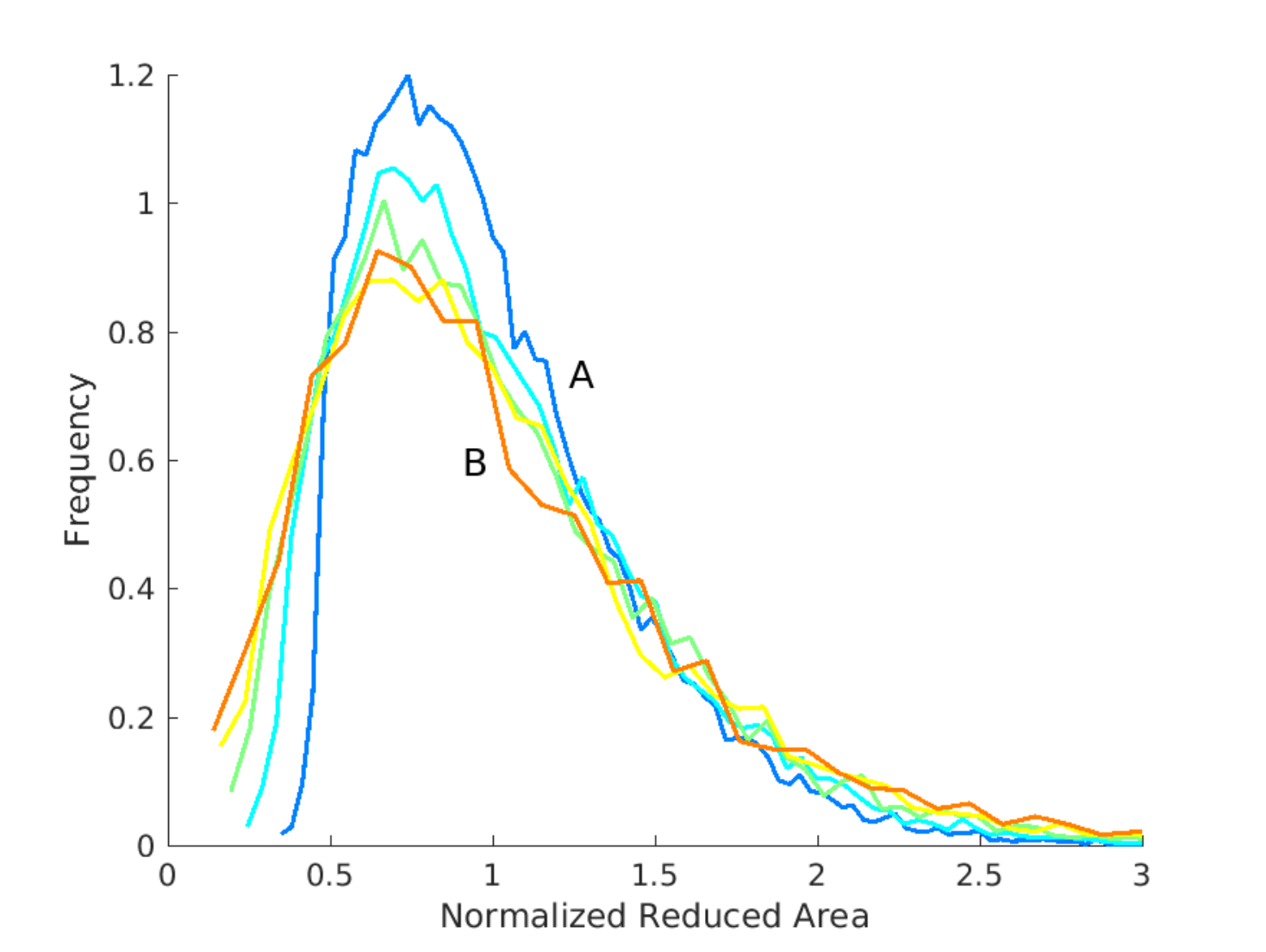}}\enskip
	\subfloat[]{\includegraphics[width=0.45\textwidth]{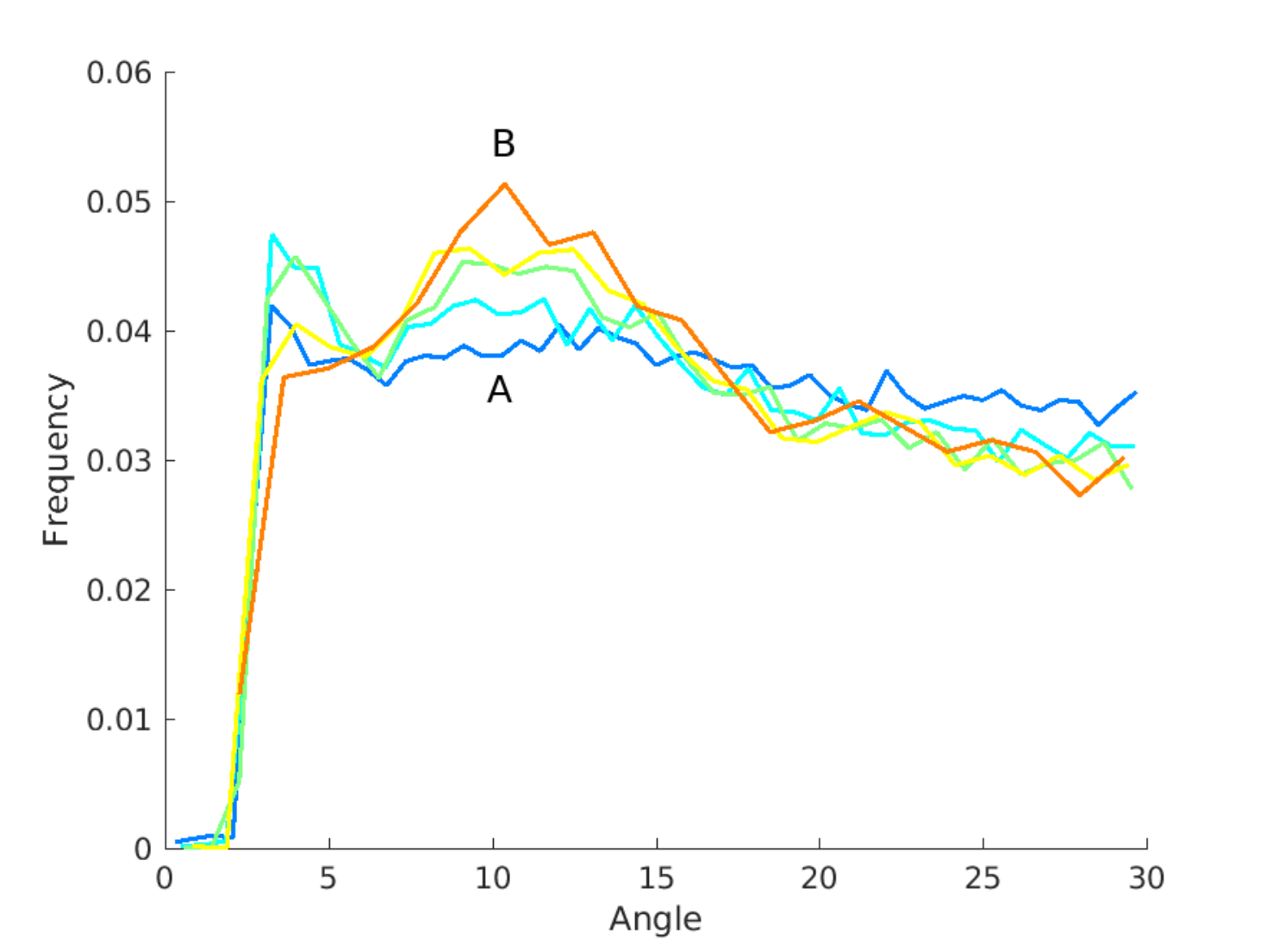}}\enskip
	\subfloat[]{\includegraphics[width=0.45\textwidth]{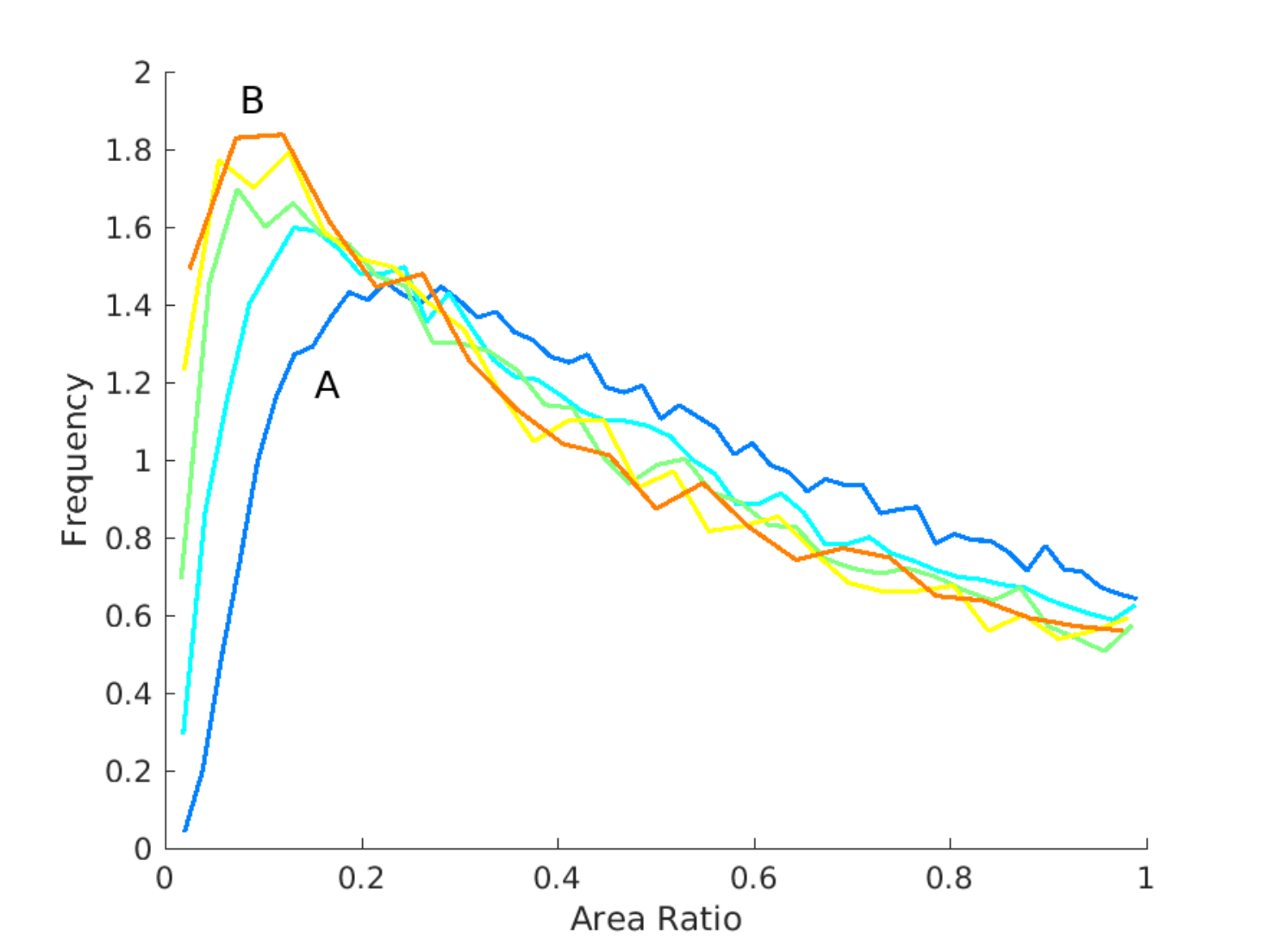}}\enskip
	\caption{Evolution in time of the normalized reduced area (a), GBCD (b) and area ratio (c) CDFs with the PFC evolution. Lines correspond to the time steps $\{228, 752, 2828, 10637, 40000 \}$ with blue, A, being early in time and orange, B, late in time. Histograms have been normalized in area to make the comparison clearer.}
	\label{fig:PFC_Geometric}
\end{figure}

Using the PFC equation, we evolved $18$ initially noisy distributions over $40000$ time steps, saving the phase field at certain intervals. Using $\theta = 2.5^\circ \midand \alpha = 40$, geometric properties were extracted and all runs were added together to form much larger data sets. Since these large sets now contain more than roughly $2300$ grains at the final time, histograms become accurate in visualizing the evolution. We present the normalized reduced area, GBCD and area ratio distributions extracted using our algorithm in figure \ref{fig:PFC_Geometric}.

The normalized reduced area behaves as in \cite{BACKOFEN_GSD}, being at least approximately lognormal with a peak becoming flatter in time, stabilizing around $0.65$. We note however that the relative proportion of very small and very large grains does continue to increase as time goes on. Perimeter essentially tracks the reduced area while the isoperimetric ratio decreases, indicating that grains become qualitatively less regular in time. The absolute orientation distribution does not evolve in time and remains constant since the PFC equation is isotropic. On the other hand, the misorientation distribution and GBCD do evolve interestingly. The GBCD tracks the misorientation distribution but its features are more pronounced as there is some additional correlation between misorientation and interface length. Early in time, the GBCD is relatively flat away from the threshold but as time goes on, the misorientations around $10^\circ$ become over represented to the detriment of large misorientations. The area ratio introduced earlier is also quite interesting. This property may be interpreted as follows: a flat area ratio distribution would indicate that the area of neighbor grains are completely uncorrelated. A peaked distribution close to $1$ would indicate the neighbor grains are likely to be similar in size while a peak close to $0$ would indicate that neighbors are likely to be very different in size. From figure \ref{fig:PFC_Geometric} (c), it is clear that the PFC evolution favors large area imbalances between neighbours. Finally, the coordination number does not evolve in time, peaking at $5$ with an average of almost exactly $6$ at all times.

\section{Conclusion}
Grain extraction being an important step in analyzing materials science data, there is a clear need for automated algorithms capable of extracting both the grain distribution and grain geometric properties. We detailed a simple, accurate and efficient atom based method to extract such grain networks along with grain area, perimeter and other properties. The detection accuracy of these measurements was tested using PFC grain distributions and surrogate artificial distributions. Overall, the boundary network and grain area may be detected with very high precision while perimeter is underestimated by less than $5\%$ for very small grains. This accuracy could be improved, for example by measuring the thickness of grain boundaries locally instead of assuming a uniform thickness for all grain sizes. Other geometric properties such as the GBCD may be computed with similar precision. We also compared the proposed method with the simpler but less accurate grid based approach.

The numerical method we have presented may then be used to automatically characterize the geometric properties of statistically significant data sets with good accuracy considering the difficulty and ambiguity inherent to the task. With modifications, our general atom based framework may be applied to 3d and to other crystal lattices. We shall use this scheme to analyze the PFC evolution in more detail in a future article.

\section*{Acknowledgments}
We are indebted to Selim Esedoglu for many useful conversations and suggestions. We are also indebted to Jean-Christophe Nave for originally suggesting an atomic Voronoi approach and for many useful conversations. GMLaB was supported by an FRQNT (Fonds de recherche du Qu\'ebec - Nature et technologies) Doctoral Student Scholarship. RC was supported by an NSERC (Natural Sciences and Engineering Research Council Canada) Discovery Grant.

\bibliographystyle{unsrt}
\bibliography{References}
\end{document}